\documentclass[structabstract]{aa}  
\usepackage{graphicx}
\usepackage{txfonts}
\usepackage{natbib}
\usepackage{lscape}
\begin{document}
   \title{Simulating polarized Galactic synchrotron emission at all
   frequencies}
 \titlerunning{Galactic synchrotron emission with the \textsc{hammurabi} code}

   \subtitle{the Hammurabi code}

   \author{A. Waelkens
          \inst{1} \fnmsep\thanks{email: waelkens@mpa-garching.mpg.de}
          ,  T. Jaffe \inst{2}, M. Reinecke \inst{1}, F. S. Kitaura \inst{1},
          T. A. En{\ss}lin \inst{1}          
          }
        \authorrunning{Waelkens et al.}

   \institute{Max Planck Institut f{\" u}r Astrophysik (MPA),
              Karl-Schwarzschildstr. 1, 85741 Garching, Germany\\
              \email{waelkens@mpa-garching.mpg.de}
         \and
             Jodrell Bank Centre for Astrophysics, University of Manchester,
              Alan Turing Building, Oxford Road, Manchester, M13 9PL, UK\\
             }

   \date{Received Month ??, ????; accepted Month ??, ????}

 
  \abstract
   {Galactic synchrotron emission, rotation measure (RM) and the deflection of
   ultra-high-energy-cosmic-rays (UHECR) permit detailed studies of the
   Galactic magnetic field (GMF). The synchrotron emission has also to be measured in order to
   be separated from other astrophysicaly interesting signals like the CMB.}
   {We present a publicly available code called \textsc{hammurabi} for generating mock polarized observations of Galactic
   synchrotron emission for telescopes like LOFAR, SKA, Planck and WMAP, based on
   model inputs for the GMF, the cosmic-ray density
   distribution and the thermal electron density. We also present mock UHECR
   deflection measure (UDM) maps based on model inputs for the GMF. In future,
   when UHECR sources are identified, this will allow us to define UDM as a
   GMF probe in a similar way as polarized radio sources permit us to define
   rotation measures.} 
 {To demonstrate the code's abilities mock observations are compared to real
 data as a means to constrain the input parameters of our simulations with a
 focus on large-scale magnetic field properties.} 
   {The Galactic magnetic field models in the literature seem to fail to
   reproduce any additional observational data which was not included in their design.}
   {As expected, attempts at trying to model the synchrotron, UHECR deflection and RM input parameters,
  show that any additional observational data set greatly increases the constraints on
  the models. The \textsc{hammurabi} code addresses this by allowing to perform simulations of several different data sets simultaneously, providing
  the means for a more reliable constraint of the magnetized inter-stellar-medium.}


             \maketitle
%

\section{Introduction}

There are several different observations which probe the Galactic
magnetic field (GMF): synchrotron radiation, rotation measure, UHECR
deflection, dust related observations (e.g. starlight polarization \citet{Heiles2000}, polarized dust emission), and more
recently atom alignment spectroscopic observations were proposed
\citep[see][]{2006ApJ...653.1292Y}.

A better understanding of the inter-stellar magnetized plasma and the radio emission
processes related to it is also paramount for current and future CMB
experiments \citep[see, e.g.,][]{MAMD2008, Page2007}, where in particular
measurements of the polarized CMB signal will be limited by our knowledge of
foreground emission \citep{Tucci2005}.

Observational studies of the GMF are usually based on one single
sort of observable. However, since the different observables and measurement
methods provide complementary information, constraining the GMF is logically
better if all possible observations are considered simultaneously.
We address the need to confront those observations with models by presenting
a publicly available software\footnote{\tt Software available at \\
    http://www.mpa-garching.mpg.de/$\sim$waelkens/hammurabi} capable
of generating mock synchrotron and Faraday rotation observations as well as
mock UHECR deflection maps. 

The code is constructed in such a fashion that it
should be feasible to adapt it to do any sort of all or partial-sky
line-of-sight integral observations, (see, e.g., \citealt{Waelkens2008} and \citealt{Sun2008}).

We will give first a brief account of previous work on GMF modeling in
Sect. \ref{sec::previous} and then we will give a description of the involved
Faraday rotation, UHECR deflection and synchrotron
physics, and the simplifying assumptions used in our computation
(Sect. \ref{sec::physics}). The implementation of a line-of-sight (hereafter
LOS) integration scheme which mimics the radiative transfer is discussed in Sect. \ref{sec::Implementation}.
We present some test output using simple standard magnetic field, thermal- and
cosmic-ray electron models. We discuss the role of the turbulent field and
sub-grid modeling in Sect. \ref{sec::SSGMF}, and the one of helical
magnetic fields in Sect. \ref{sec::HGMF}. Finally we conclude in Sect. \ref{sec::Conclusion}.

\section{Previous work on Galactic magnetic field modeling}
\label{sec::previous}
GMF modeling already counts with several contributions from the community. See,
e.g., \citet{Beuermann1985,Han2006,Brown2003,Haverkorn2006,Page2007,Sun2008} and
many others.
The data used are mostly a number of total and polarized emission radio
surveys of the Galaxy \citep{CGS2003, Gaensler2001, Haslam1982, Reich1986, Page2007} as
well as measurements of Faraday rotation \citep{Dineen2005, JH2004, Han2006} and a starlight
polarization catalogue compiled by \citet{Heiles2000}.
Theoretical predictions are also heavily based on experience obtained from synchrotron observations of other
spiral Galaxies \citep[see e.g.][]{Beck1996}.
Our work complements previous efforts by making it possible to compare simultaneously the largest
possible number of observables to theoretical predictions.


\section{Physics included in the code}
\label{sec::physics}
Here we describe the physical processes included and the assumptions underlying
the code.
\subsection{Faraday rotation}
The polarization angle of an electromagnetic wave is rotated when
crossing a magnetized plasma. This effect is known as Faraday rotation (see
e.g. \cite{RL} ). The observed polarization angle $\chi$ will be a function
of the rotation undergone when crossing the magnetized plasma, the square of
the observation wavelength and its original
(or intrinsic) angle $\chi_0$ at the polarized source, 
 \begin{equation}
   \label{eq::FR}
   \chi = RM \, \lambda^2+\chi_0 \, .
 \end{equation}
The rotation measure (hereafter RM), which quantifies the linear rate of change
of the angle $\chi$ as a function of $\lambda^2$, is a
function of the integral of the magnetic field $\vec{B}_{LOS}$ along the LOS weighted by the thermal electron density $n_e$,
\begin{equation}
\label{eq::RM}
RM=a_0 \int_{here}^{there} dr \, n_e \,
B_{LOS} \, ,
\end{equation}
where $a_0=q_e^3/(2\pi m_e^2c^4)$, $m_e$ is the electron mass, $c$ is
the speed of light, and $q_e$ is the electron charge.

The RM can be measured directly via a fit to Eq. \ref{eq::FR} only in the particular case of a Faraday
screen, i.e., if the observer and the polarized source have a cloud of magnetized
plasma in between them, but no source-intrinsic Faraday rotation occurs. 
In the more complicated scenario of several sources along the LOS embedded in the magnetized plasma, the RM cannot be
measured in that way since the polarization angle will no longer obey a
linear dependence on $\lambda^2$. A complicated frequency dependence of the
polarization angles arises in such a case, and that is the situation which is
typically found in our Galaxy. There, the synchrotron-emitting cosmic-ray electron
population is embedded in the magnetized plasma that is producing the Faraday
rotation simultaneously. 

\subsection{Synchrotron emission, total and polarized}
\label{sec::SyncRad}
Relativistically moving charges in a magnetic field emit
synchrotron radiation. In the Galactic case we deal with a cosmic-ray electron
population mostly arising from supernova explosions
and subsequent shock acceleration, and the GMF of the order
of a few micro Gauss with a yet unknown topology.
In the following, a couple of simplifying assumptions are made:

\begin{itemize}
\item The relativistic CR electrons have an isotropic velocity distribution, as
  is measured with high accuracy to be the case at our location in the
  Galaxy (their propagation, though, is not isotropic due to the presence of
  ordered magnetic fields; \citealt{Yan2008}). 
\item The cosmic-ray electron spectrum is assumed to be a power law with
  spectral index $p$. This widely used simplification is motivated by the
  theory of shock acceleration (a.k.a. Fermi acceleration,
  which predicts power law energy distributions; \citealt{Drury1983}) and is simultaneously
  confirmed by the measured cosmic-ray electron spectrum at Earth
  \citep[e.g.][and references therein]{CRE.earth}. The same observations, as
  well as sophisticated simulations have, however, shown that the power
  law assumption is not adequate for the entire energy spectrum \citep[see][]{Strong2007}.
\end{itemize}

The synchrotron emissivity (i.e. power per unit volume per frequency per solid
angle) is partially linearly polarized. Its intensity and polarization
properties depend on the strength and orientation of the
perpendicular (with respect to the
LOS) component of the magnetic field, $B_\perp$, and the cosmic-ray electron spatial and energetic distribution. The
emissivities are usually subdivided into two components,
$j_{\perp,\parallel}=dE_{\perp,\parallel}/dt \, d\omega \, d\Omega \, dV$, respectively
perpendicular and parallel to $B_{LOS}$, following \citet{RL} and \citet{1959ApJ...130..241W}:
\begin{eqnarray}
j_{\perp} (\omega, \vec{r}) & = & \frac{1}{4\pi}
\frac{\sqrt{3}q^3}{8 \pi mc^2} \omega^{\frac{1-p}{2}}  \left
  (\frac{2mc}{3q} \right )^{\frac{1-p}{2}} {B_\perp (\vec{r})}^{\frac{p+1}{2}} C(\vec{r}) \nonumber \\
&& \!\!\!\!\!\!\!\!\!\!\!\!\!\!\!\!\!\!\!\!\!\!\!\!\! \times \, \Gamma \left ( \frac{p}{4}-\frac{1}{12} \right) \left
  [\frac{2^{\frac{p+1}{2}}}{p+1} \Gamma \left ( \frac{p}{4}+\frac{19}{12}
  \right) + 2^{\frac{p-3}{2}} \Gamma \left ( \frac{p}{4}+\frac{7}{12} \right)
\right ] \, , \label{eq::jPerp}
\end{eqnarray}
and
\begin{eqnarray}
j_{\parallel} (\omega, \vec{r}) & = & \frac{1}{4\pi}
\frac{\sqrt{3}q^3}{8 \pi mc^2} \omega^{\frac{1-p}{2}} \left (
  \frac{2mc}{3q} \right )^{\frac{1-p}{2}} {B_\perp
(\vec{r})}^{\frac{p+1}{2}} C(\vec{r}) \nonumber \\
&& \!\!\!\!\!\!\!\!\!\!\!\!\!\!\!\!\!\!\!\!\!\!\!\!\! \times \, \Gamma \left ( \frac{p}{4}-\frac{1}{12} \right) \left
  [\frac{2^{\frac{p+1}{2}}}{p+1} \Gamma \left ( \frac{p}{4}+\frac{19}{12}
  \right) - 2^{\frac{p-3}{2}} \Gamma \left ( \frac{p}{4}+\frac{7}{12} \right)
\right ] \, . \label{eq::jPar}
\end{eqnarray}
Here $C$ depends on the position in the Galaxy  and is defined by $N(\gamma) d\gamma= C \gamma^{-p} d\gamma$,
$\gamma$ being the Lorenz factor, $N(\gamma)$ the number density of electrons $\in
[\gamma, \gamma+d\gamma]$, and $p$ is the spectral index as mentioned above. The charge of the electron is given by $q_e$, the mass by
$m_e$ and $\omega=2 \pi \nu$, where $\nu$ is the observation frequency.
The specific intensity ${\it I}$ as a function of observation frequency and
LOS direction $\vec{\hat n}$ is
\begin{equation}
\label{syncTeo:speI}
\it {I}(\omega,\vec{\hat n})=\int_0^{\infty} dr \: \left [ {j_\perp(\omega, r
    \vec{\hat n})}+{j_\parallel(\omega,r \vec{\hat n}} \right ],
\end{equation} and the polarized specific intensity ${\it P}$ expressed as a complex variable
is \citep[see][]{1966MNRAS.133...67B}:
\begin{eqnarray}
{\it {P}(\omega,\vec{\hat n})}&=&\int_0^{\infty} dr \: \left (j_{\perp}(\omega,r \vec{\hat n}) \right . \nonumber  \left .-{j_{\parallel}(\omega,r \vec{\hat n})} \right )e^{-2 i \chi(r \vec{\hat n})} .
\label{syncTeo:P}
\end{eqnarray}
The intrinsic emission polarization angle $\chi_0$ is given by the inclination
towards the Galactic North Pole of the local perpendicular-to-the-LOS component
of the magnetic field at each position in space \citep[same convention as
adopted in][]{Page2007}. 
The Stokes {I}, {Q}, {U} parameters \footnote{Here the specific intensities are in
  italic ({\it I},{\it Q},{\it U}), while the observed Stokes parameters
  (I,Q,U) are denoted by Roman letters.} are then the integrals over solid angle $\Omega$:
\begin{eqnarray*}
{\rm I}=\int d\Omega \, \it {I},
\end{eqnarray*}
and
\begin{eqnarray*}
{\rm Q}+i {\rm U} = \int d\Omega \, \it {P} .
\end{eqnarray*}

\subsection{UHECR propagation}
UHECR's are deflected by the GMF due to the Lorentz force. The Larmour radius
for relativistic particles is $r_g = p\, c/ Z \, q_e \, B_\perp$, $Z\, q_e$ being the charge of the
UHECR, and p the momentum perpendicular to the magnetic field $B_\perp$.
Provided, the sources of UHECRs could be identified, an UHECR deflection
measure (UDM) can be extracted from the UHECR arrival distribution via fitting
the arrival data (position \& energy).
The net deflection of an UHECR can be approximated by the LOS integral
\citep[see e.g.][]{Kach2007}
\begin{eqnarray}
\label{eq::UHECRDA}
\Theta_{\rm offset} \approx \int dl \, \vec r_g^{-1} = \frac{Z \, q_e}{pc}\int dl \vec
B_\perp \,  \equiv \frac{Z \, q_e}{pc} \vec {UDM}.
\end{eqnarray}

\section {Implementation}
\label{sec::Implementation}
In this section we present the implementation of the physics described in
Sect. \ref{sec::SyncRad} and the technical characteristics of the \textsc{hammurabi} code.
Given a
\begin{itemize}
\item 3D GMF model,
\item 3D cosmic-ray electron density model,
\item 3D thermal electron density model,
\end{itemize}
\textsc{hammurabi} computes full sky maps for
\begin{itemize}
\item the Galactic RM contribution to the extra-galactic sources, 
\item synchrotron I,Q and U Stokes parameters, taking into account intrinsic
  Galactic Faraday depolarization and,
\item UHECR deflection intensity and orientation maps.
\end{itemize}

The sky maps in \textsc{hammurabi} are subdivided into equal area pixels following
the HEALPix\footnote{http://healpix.jpl.nasa.gov} pixelization scheme of
\cite{2005ApJ...622..759G}. The total number of pixels for an all-sky map,
which defines the angular resolution, is $N_{\rm pix}=12{\rm NSIDE}^2$, with
${\rm NSIDE}=2^k$ and $k=0,\ldots,13$ (a HEALPix package limitation, which
  can be altered, see App. \ref{app::HPG}).  The angular size of a pixel
($\Delta\theta$) can be approximated as  
\begin{equation}
\Delta\theta\approx\sqrt{\frac{3}{\pi}}\frac{3600\arcmin}{\rm NSIDE} \, .
\end{equation}

The observation volume associated with one of the HEALPix-map pixels is sampled along the LOS at a constant
radial interval $\Delta r$. As a consequence of the conical shape of the
 implied effective observation beam, the volume units increase with radius, and hence the weights of the sampling
points along the LOS do also increase. To limit the
amount of non-homogeneity of the sampling, the code allows the
volume resolution to be increased by splitting the beam in sub-beams at some
radius, which subsequently can be further split later on. We call the implied 3-D sampling grid
the ``3D HEALPix grid''. See further details in App. \ref{app::HPG}.

Formally, the maximally achievable volume resolution, i.e. the largest volume unit at the
finest possible resolution $\Delta \theta \sim 0.43 \arcmin$, (corresponding to
${\rm NSIDE}=2^{13}$) is $V_{ext}=4\pi R_{max}^2 \Delta
r(12\cdot 2^{26})^{-1} \sim \left (4 {\rm pc} \right)^3$ for $R_{max} \sim 32
{\rm kpc}$ and $\Delta r = 4 {\rm pc}$ since we impose an approximately cubic volume unit according
to $\Delta r \sim (4 \pi R_{max}^2/12 \cdot 2^{26})^{1/2}$. This implies that
variations of the input parameters (magnetic field, thermal electron
distribution,...) on scales smaller than this volume can only be taken into
account with sub-grid modeling if they persist to the largest radii simulated. Note however that this is an upper limit,
given that fluctuations are likely to be stronger closer to the Galactic
disk, where resolution, due to the cone-like shape of the observation volume,
is going to be higher anyways.

The integration is performed assuming an optically thin medium. \citet{Sun2008}
enhanced the code by introducing free-free absorption. This implementation,
relevant mainly at frequencies below a GHz, is present in the code but not
described here. Furthermore, \citet{Sun2008} also introduced a coupling between
the thermal electron density and the random component of the RM. This
implementation is necessary to explain the degree of depolarization at their
simulated 1.4GHz map. It is also present in the code but not described
here. For further details see \citet{Sun2008}.

The 3D HEALPix grid, since the value of one observation pixel stems from the
contribution of several sub-beams, allows one approximately to take into account
effects like beam depolarization. See App. \ref{app::LOSI} for details.

\subsection{Features}
\textsc{hammurabi} is also suited for simulations of partial sky
coverage. A single pixel or a list of pixels representing either patches or separate locations
on the sky output maps can be computed.
An extension to that is the option to compute RMs for an individual pixel only up to a certain
specified distance. This is relevant for simulating RM observations of radio
pulsars or other polarized sources with distance information in
our own Galaxy. 
Polarized sources along the same LOS allow us to probe fractions of the ISM
plasma, unlike extra-galactic polarized sources which give us the integrated RM
along the entire LOS through the Galaxy.
The mock observations for this case are done without beam-splitting and at
highest possible angular resolution, since RM are effectively obtained from
point sources.

\section{Examples}
In this section we choose standard 3D input models for a demonstration of the
code's results. It is not the scope of this work to present new scientific
findings, but to present a proof of concept of what can be done with
\textsc{hammurabi}. An actual application of the code has been done in
\citet{Sun2008}, and we will briefly refer to some of their results here. Their
preferred GMF model and cosmic-ray electron model, derived by trying to fit a
broad range of observables, are compared to the corresponding models suggested
by \citet{Page2007}, fitted to reproduce solely the polarization
angles observed by the WMAP satellite. Note that it has been shown by \citet{Sun2008}
that the \citet{Page2007} model, as well as every model from the literature
analyzed in their work, fail to reproduce different sorts of observations of
the magnetized ISM satisfactorily, since they are all constructed by
considering only a particular sort of data. The discrepancies of the
models presented here, visible by eye, are suitable for displaying the codes abilities.

\subsection{Inputs}
\subsubsection{The thermal electron density model}
For the thermal electron distribution we use the NE2001 model \cite[see
][]{CL2002aph}. This model subdivides the Galaxy into several large-scale
structure elements like a thin disk, a thick disk, spiral arms, as well as
some local small-scale elements such as supernova bubbles. 
\subsubsection{The cosmic-ray electron density model}
\label{sec::CREmodel}
We use the cosmic-ray electron density models suggested by
\citet{Page2007} and \citet{Sun2008}. 
\begin{itemize}
\item The model in \citet{Page2007} consists of an exponentially decaying disc with
characteristic hight $h_d=1 \, {\rm kpc}$ and a characteristic radius of $h_r =
5 \,
{\rm kpc}$.
Note that to compute the synchrotron emissivity, as can be seen
in Eqs. \ref{eq::jPerp} and \ref{eq::jPar}, we need the spatially dependent function
$C$, not the cosmic-ray electron density. However, since we are assuming a
Galaxy-wide unique power-law energy spectral slope with a spectral index $p=3$
(note that in principle the code allows to associate a different spectral index for each volume unit), these
quantities are proportional to each other. The value of $C_{\rm Earth}= 6.4 \cdot 10^{-5} {\rm
  cm}^{-3}$ at Earth's position is observed \citep[see e.g. Fig. 4
of][]{Strong2007}, and although it is not clear that it is representative for
other regions in the Galaxy \citep{Sun2008, Pohl1998, Strong2004}, we use it as
a zeroth order approximation for the normalization (which is not
necessary/provided in \citet{Page2007}) of our distribution, 
\begin{eqnarray}
C = C_0 \, exp\left[-r/h_r \right]\, sech^2 {(z/h_d)} \, .
\end{eqnarray}
Here $r$ is the Galactic radius, while $z$ is the hight. $C_0$ is such that
$C=C_{\rm Earth}$ at Earths position.
\item \citet{Sun2008} proposes 
\begin{equation}
C(R,z)=C_0\exp\left(-\frac{R-R_\odot}{8\,\,{\rm kpc}}-\frac{|z|}
{1\,\,{\rm kpc}}\right)
\end{equation}
with $C_0=C_{\rm Earth}$, while $C(R<3\,\,{\rm kpc})=C(R=3\,\,{\rm
  kpc})$ and $C(|z|>1)=0$. The abrupt truncation at a scale hight of $|z|>1$ as
  discussed by \citet{Sun2008} is necessary to acomodate a low
  synchrotron emission at high latitudes where an anomalously strong halo
  magnetic field is required to account for high RM
  measurements. \citet{Sun2008} warn that this seems unrealistic and suggests
  that a larger scale hight of the thermal electron density could resolve the
  problem by diminishing the required value of the halo magnetic field. They
  conclude that the high-latitude thermal electron density should be better investigated.
Furthermore the spectral index $p$ is 3 for observation frequencies larger
  than 408 MHz and 2 otherwise. \citet{Sun2008} adopt this simplification based
  on observations of \citet{Roger1999} and \citet{Reich1988a, Reich1988b}, which support a flatter
  spectrum below 408 MHz.
\end{itemize}

As mentioned in Sect. \ref{sec::SyncRad}, simulations \citep[see the GALPROP
code by][]{Strong2007} show that the assumption of a power-law energy
distribution is not applicable to the entire spectrum. The same simulations
also present a seizably more sophisticated cosmic-ray electron
distribution. 
Incorporating the output of those simulations into \textsc{hammurabi} is planned as
an extension.

\subsubsection{Large-scale Galactic magnetic field models}
The behavior of the synchrotron emissivity is mainly driven by the magnetic field distribution, as shown in Eqs. \ref{eq::jPerp} and
\ref{eq::jPar}.

It is a common practice to subdivide the Galactic magnetic field in a large-scale and
a small-scale component. The latter will be addressed in the next section
\ref{sec::SSGMF}. The subdivision in these two classes of fields is somewhat
arbitrary, and we adopt here the convention that the large-scale field is
statistically anisotropic at any point in the galaxy, while the small-scale
field is not. 

GMF modeling has been done in direct studies of the galactic
magnetic field. See \citet{Page2007,Sun2008, Jansson2007, Haverkorn2006,Brown2003,Han2006},
and others. In the context of UHECR propagation \citep[see][]{Kach2007,TT2002,PS2003,Harari2000}. 
Typically GMF models, inspired by observations of
the radio-polarization patterns of other spiral galaxies, present spiral-like structures. Here we present
simulations using the model from \citet{Page2007} and the favored (ASS+RING) model proposed by \citet{Sun2008}. The expressions are reproduced in the
App. \ref{app::MFM}. Further magnetic field models can easily be
incorporated in the code. 

\subsubsection{Small-scale magnetic field} 
\label{sec::SSGMF}
Although the turbulent component of the GMF has been studied extensively
\citep[see e.g.][]{Haverkorn2003, Han2004, Haverkorn2006, Haverkorn2008} there is, to
our knowledge, not yet any model which describes it to a satisfactory
degree. Observational constraints on its relative strength compared to that of
the large-scale field put it at roughly the same amplitude (i.e. a couple of
micro Gauss; \citealt{Beck2001}).
\textsc{hammurabi} allows one to simulate, given a magnetic field
power spectrum, a Gaussian random field realization. A realistic turbulent
field, however, needs not be Gaussian, since simulations of turbulence show
sheet like structures evidently different from what is expected from a Gaussian
field \citep[see e.g.][]{2004ApJ...612..276S, Schekochihin2006}.
This means that the Gaussian random field should per construction reproduce the
correct two point statistics of a real field, however it will probably not
reproduce higher order statistics as real fields are not observed to be
Gaussian \citep[see e.g. the figures in ][]{Jaegers1987,Clarke2006}. Furthermore, this modeling assumes the power spectrum of the
field to be known, which in the case of the Galaxy, to our knowledge, has not
yet been determined. There are, however, measurements of the magnetic power
spectrum in the intra-cluster medium \citep [e.g.][]{Vogt2005, Guidetti2008,
  Govoni2006}. 

\subsubsection{Helical fields}
\label{sec::HGMF}
Helical magnetic fields have a broken mirror symmetry and can best be visualized
by twisted flux tubes. Topological helicity is given by $\int \, dV\, A\cdot
B$, where A is the vector potential of the field, $B=\nabla \times A$. Current
helicity is given by $\int \, dV\, j\cdot B$, where j is the electrical current,
satisfying $j=\nabla \times B$. Both types of helicity are closely
related. Topological helicity is a nearly conserved quantity, even in non-ideal
MHD, and plays a crucial role in magnetic dynamo theory. Therefore \textsc{hammurabi} was adapted to read in random small-scale fields with
magnetic helicity generated by a separate code. This code, also available
for download, is called \textsc{garfields} \citep[first applied
in][]{Kitaura2007}, where helicity is imprinted onto a divergence free Gaussian
random field $\vec{B}'$ by
\begin{equation}
\hat B_m=\frac{1}{\sqrt{1+\alpha^2}}\left ( \delta_{jm}+ i \alpha
  \epsilon_{ijm} \overline k_i\right
)\hat B'_j .
\end{equation}
Here $\hat{B}'_i=\hat{B}'_i(\vec{k})$ and it is such that $\langle \hat{B}'_i
  \hat {B}'^*_j \rangle = \left (  \delta_{ij}-\overline k_i
  \overline k_j \right )P(k)/2$, where $P(k)$ is the magnetic power spectrum. The hat denotes a Fourier transform, $\delta$ is the Kronecker
delta, $\overline k_i$ is an unit vector and $\epsilon_{ijk}$ is the
permutation symbol. Maximal helicity is found for $\alpha=-1$ and $1$. This leads to a magnetic
correlation tensor of the form
\begin{eqnarray}
\langle \hat{B}_i \hat{B}^*_j \rangle = \left
[ \frac{1}{2} \left(\delta_{ij}-\overline k_i \overline k_j \right)  - i
\epsilon_{ijk} \overline k_k
\frac{\alpha}{1+\alpha^2} \right] P(k) .
\end{eqnarray}

\subsection{Synchrotron and RM outputs: comparison with observations}
\begin{figure*}
\begin{tabular}{c c c c}        
\hline\hline                 
{\bf Page et al. 2007} & {\bf Sun et al. 2008} & {\bf observations  }\\    
\hline                        
\hline                                   
  \includegraphics[angle=+90, width=0.31\textwidth]{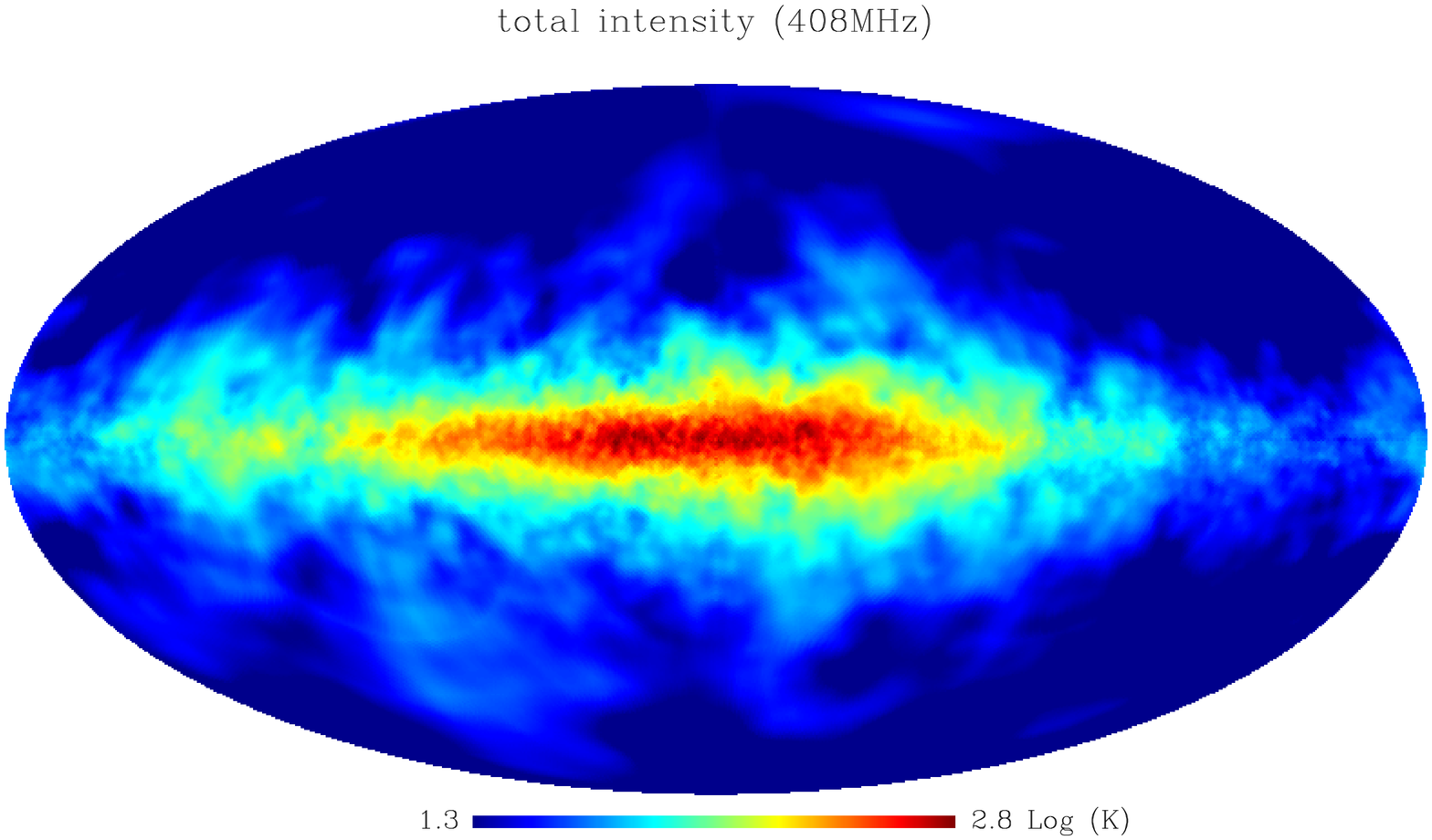} &
  \includegraphics[angle=+90, width=0.31\textwidth]{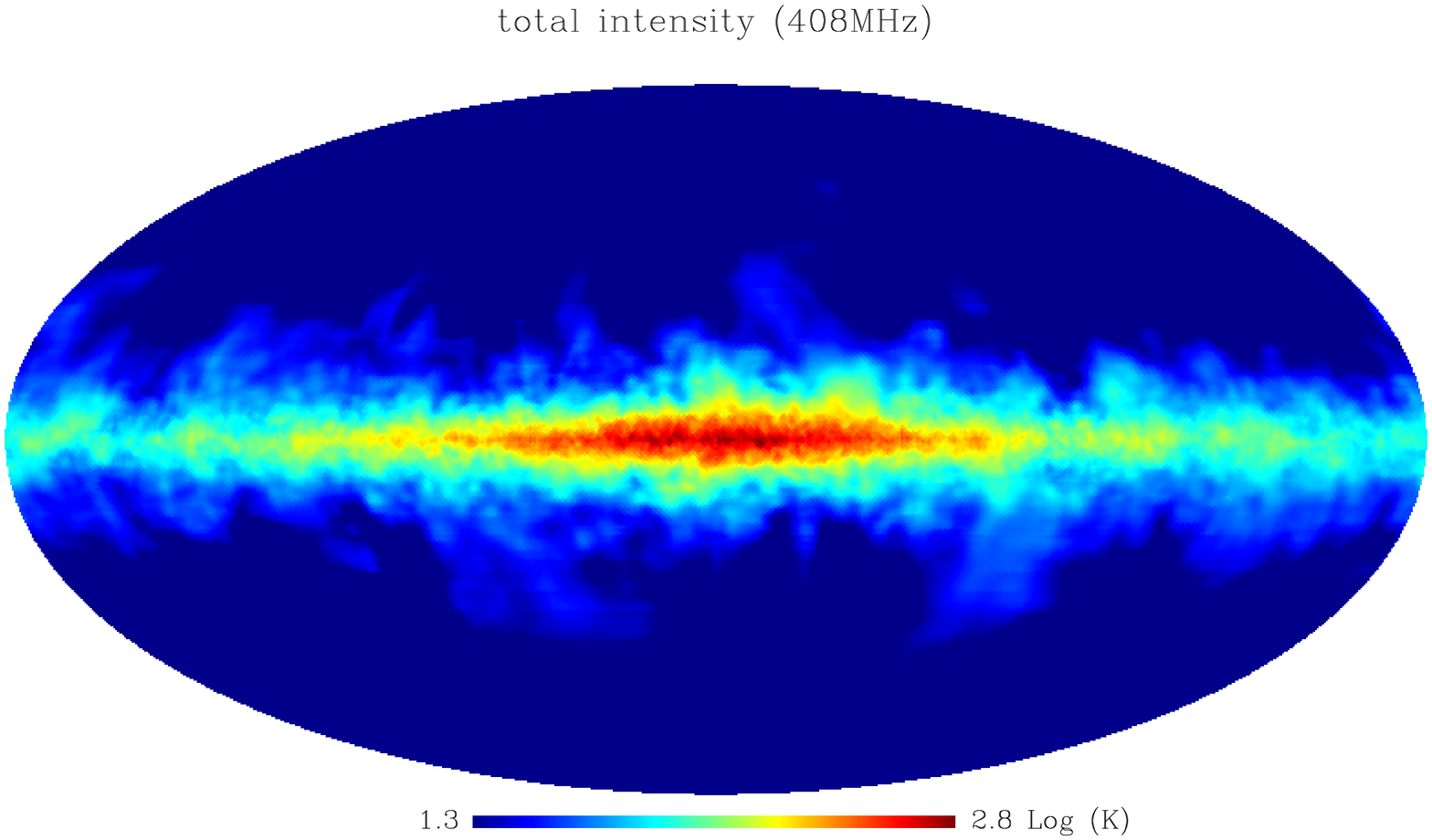}&
  \includegraphics[angle=+90, width=0.31\textwidth]{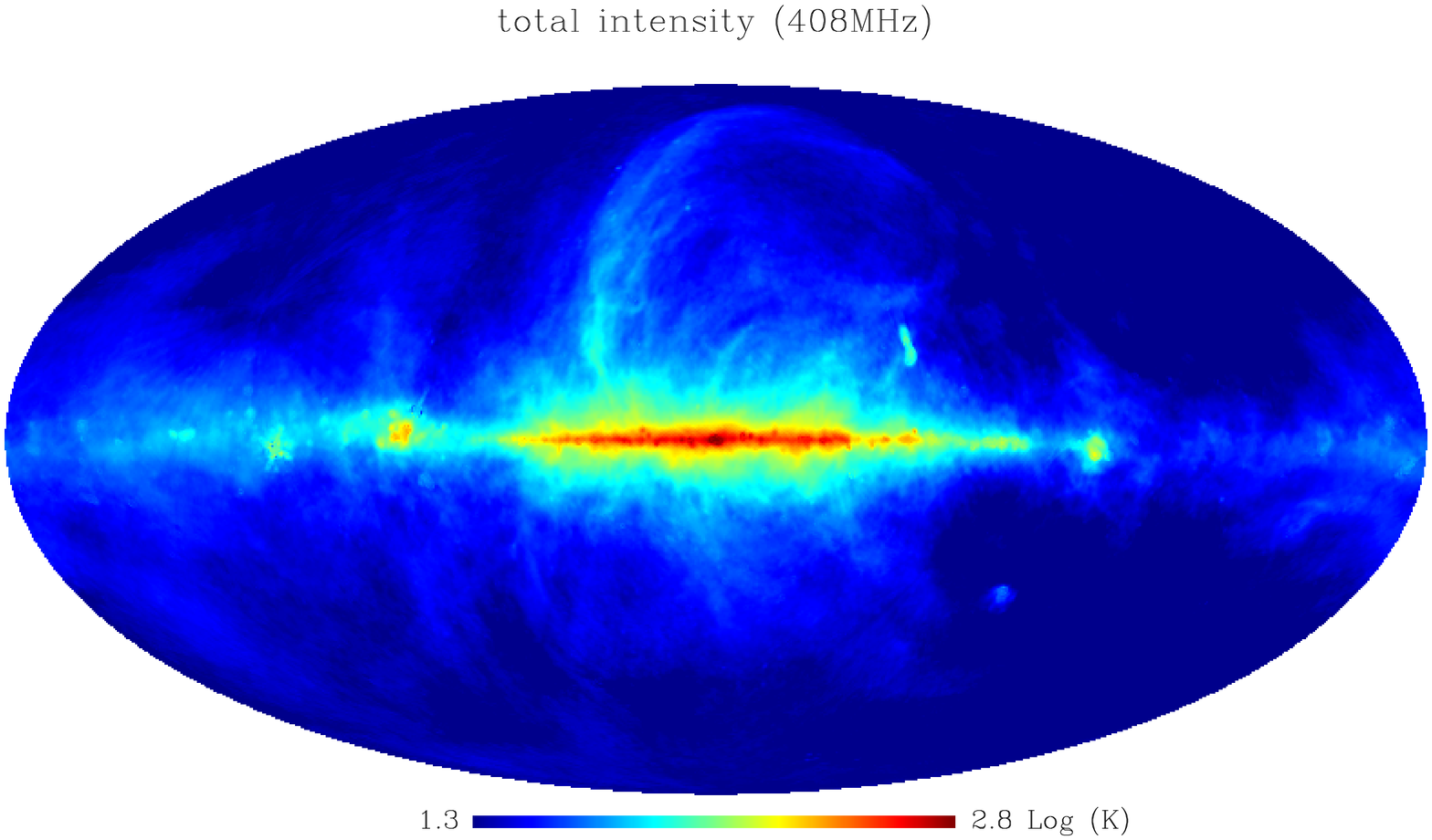} \\
  \includegraphics[angle=+90, width=0.31\textwidth]{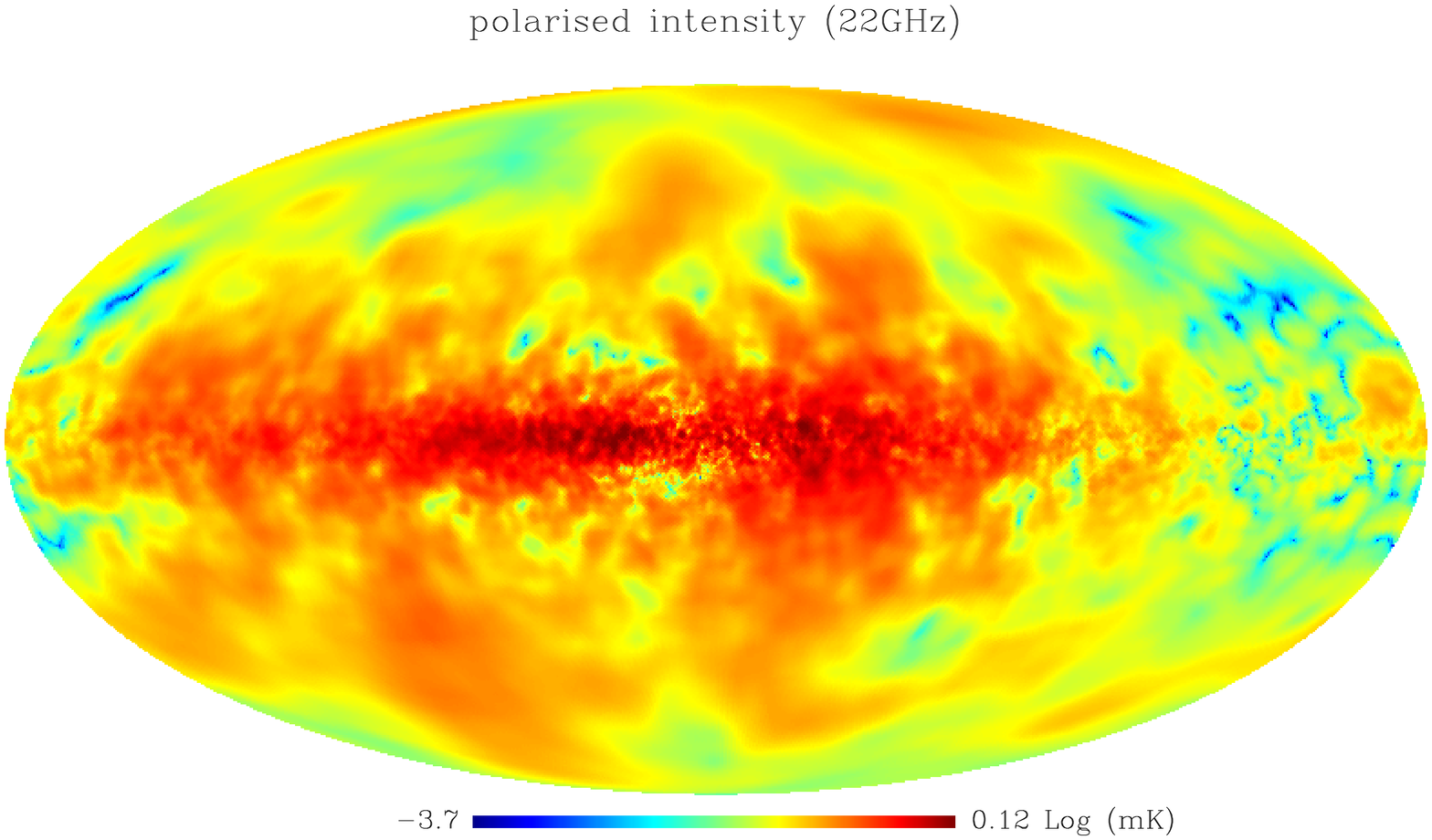}&
  \includegraphics[angle=+90, width=0.31\textwidth]{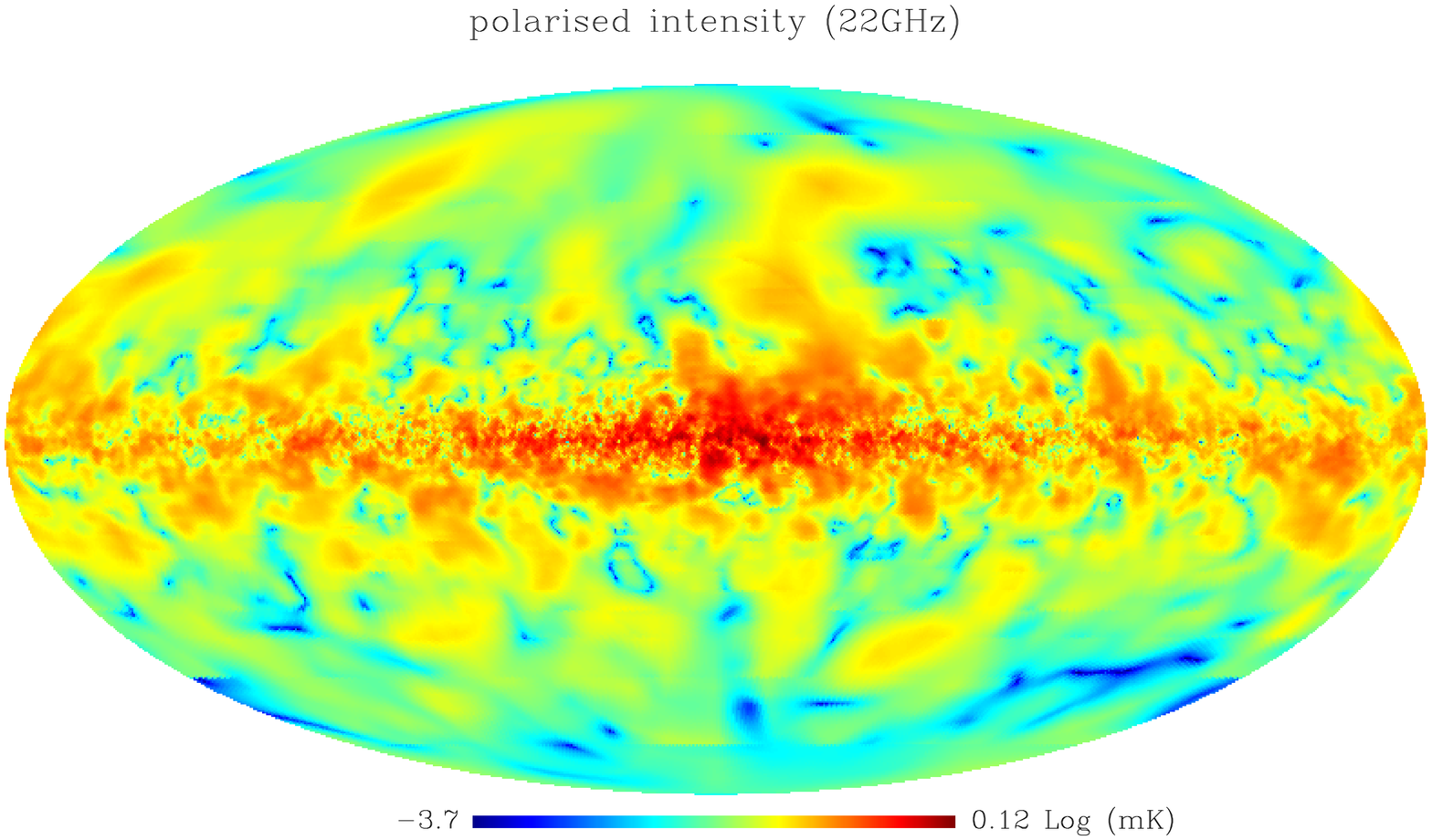}&
  \includegraphics[angle=+90, width=0.31\textwidth]{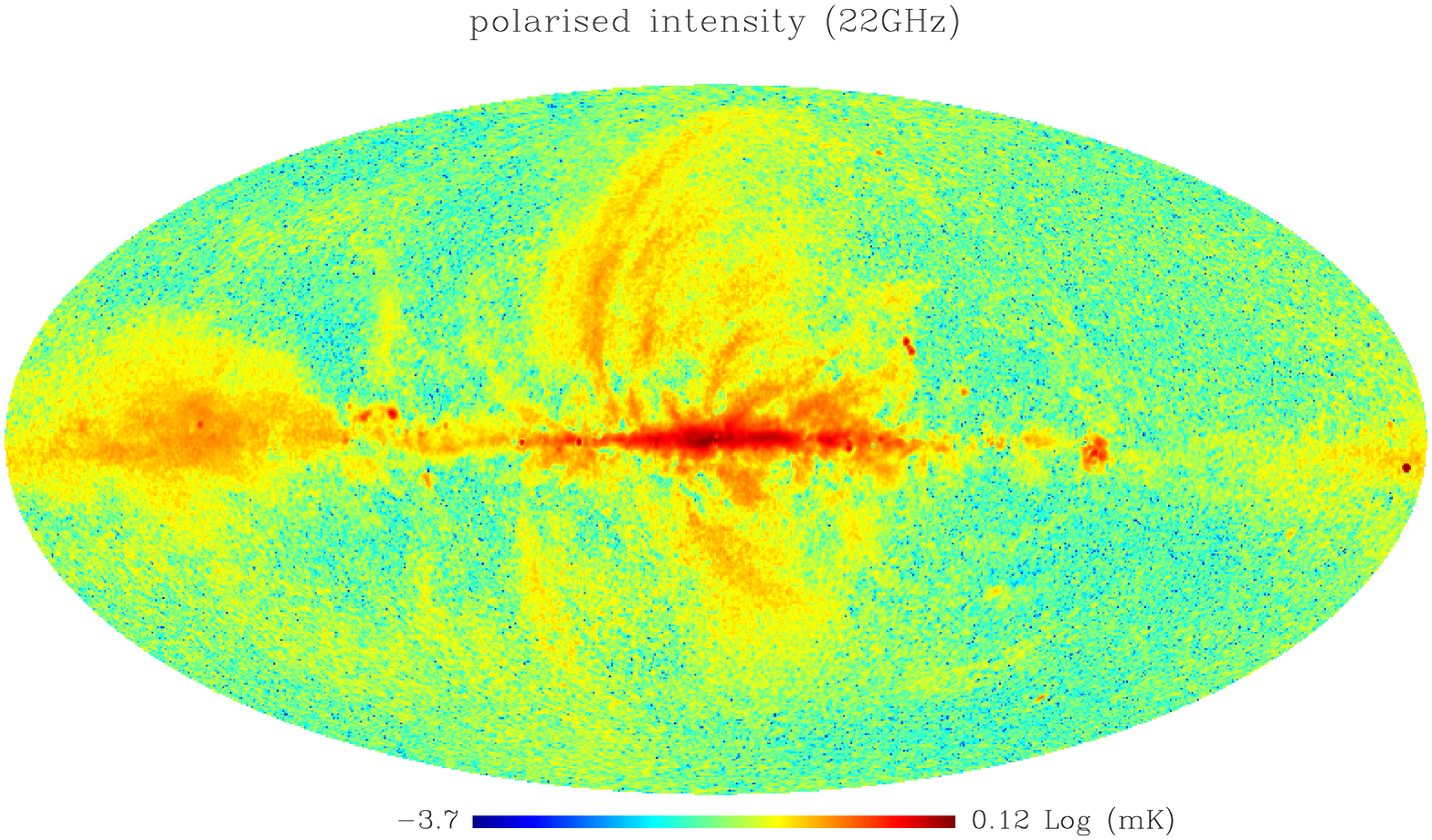}\\
  \includegraphics[angle=+90, width=0.31\textwidth]{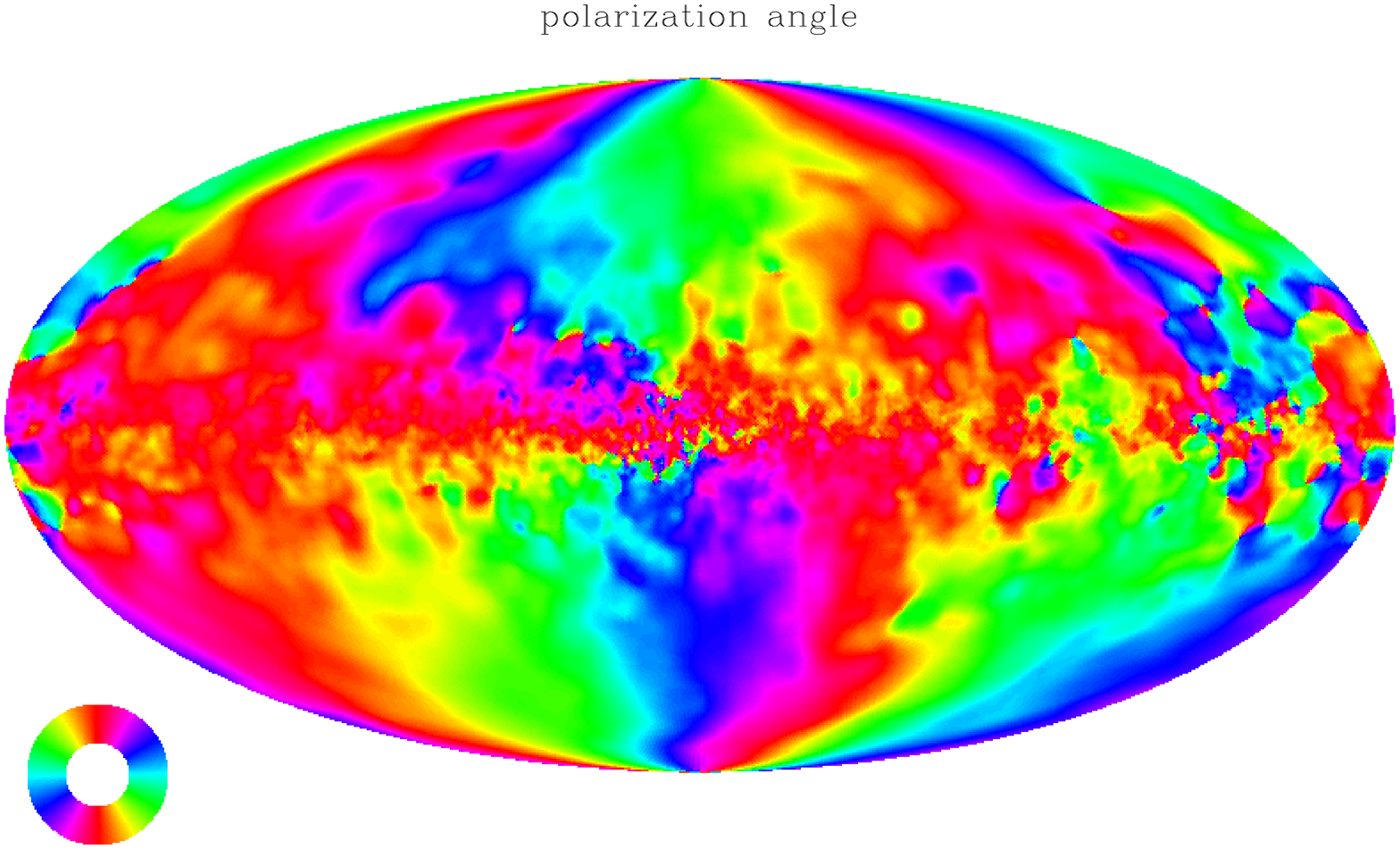}&
  \includegraphics[angle=+90, width=0.31\textwidth]{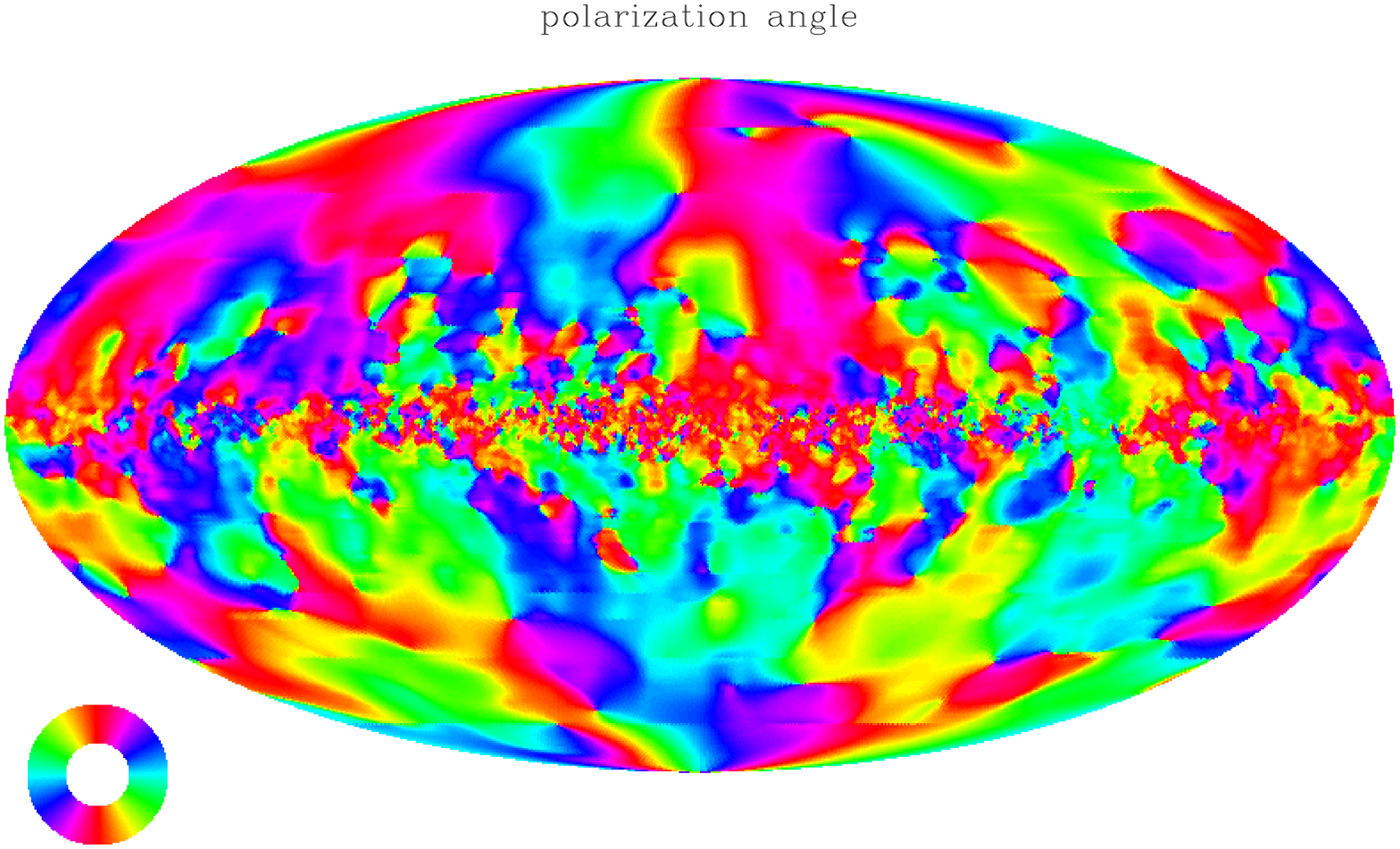}&
  \includegraphics[angle=+90, width=0.31\textwidth]{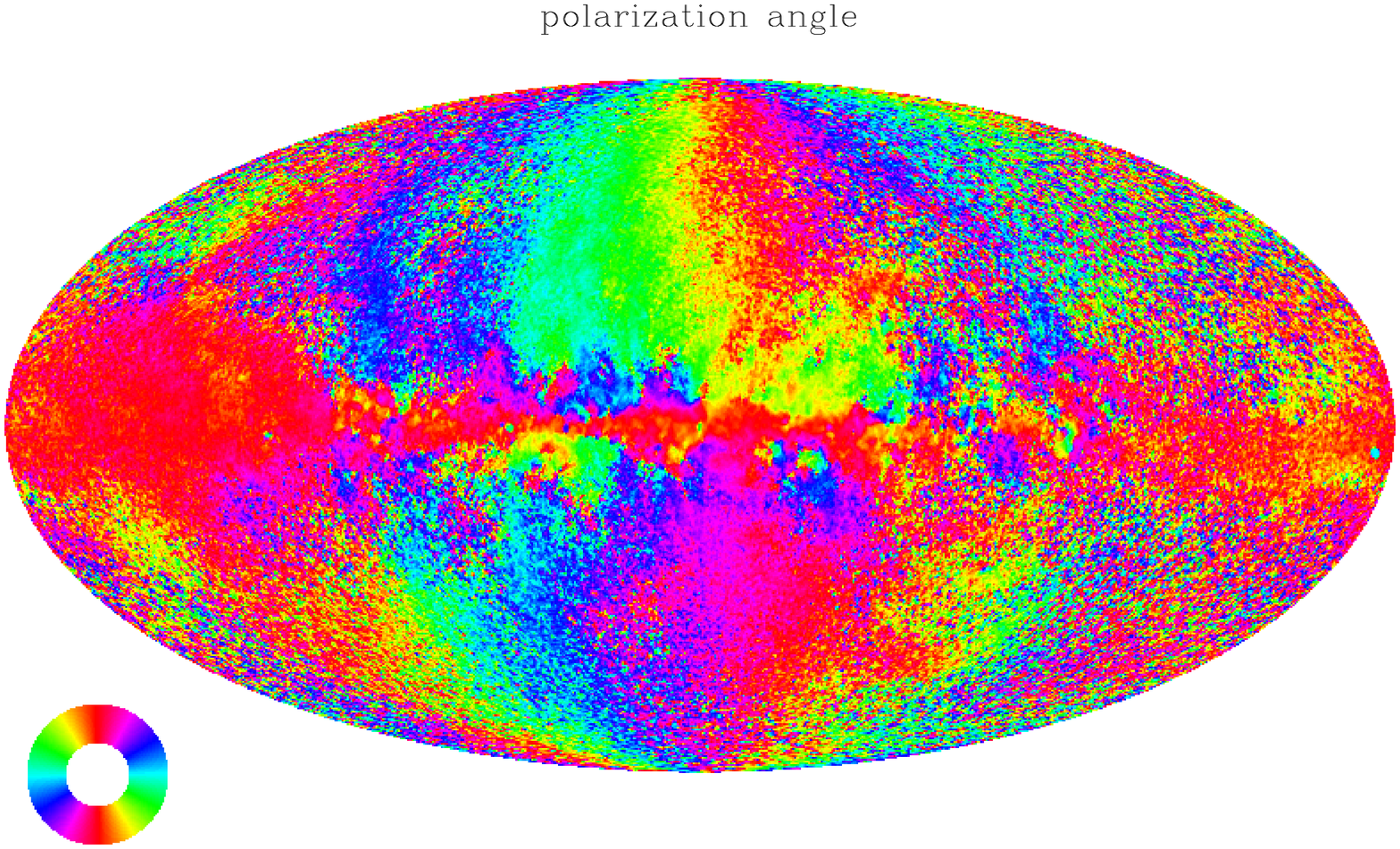}\\
  \includegraphics[angle=+90, width=0.31\textwidth]{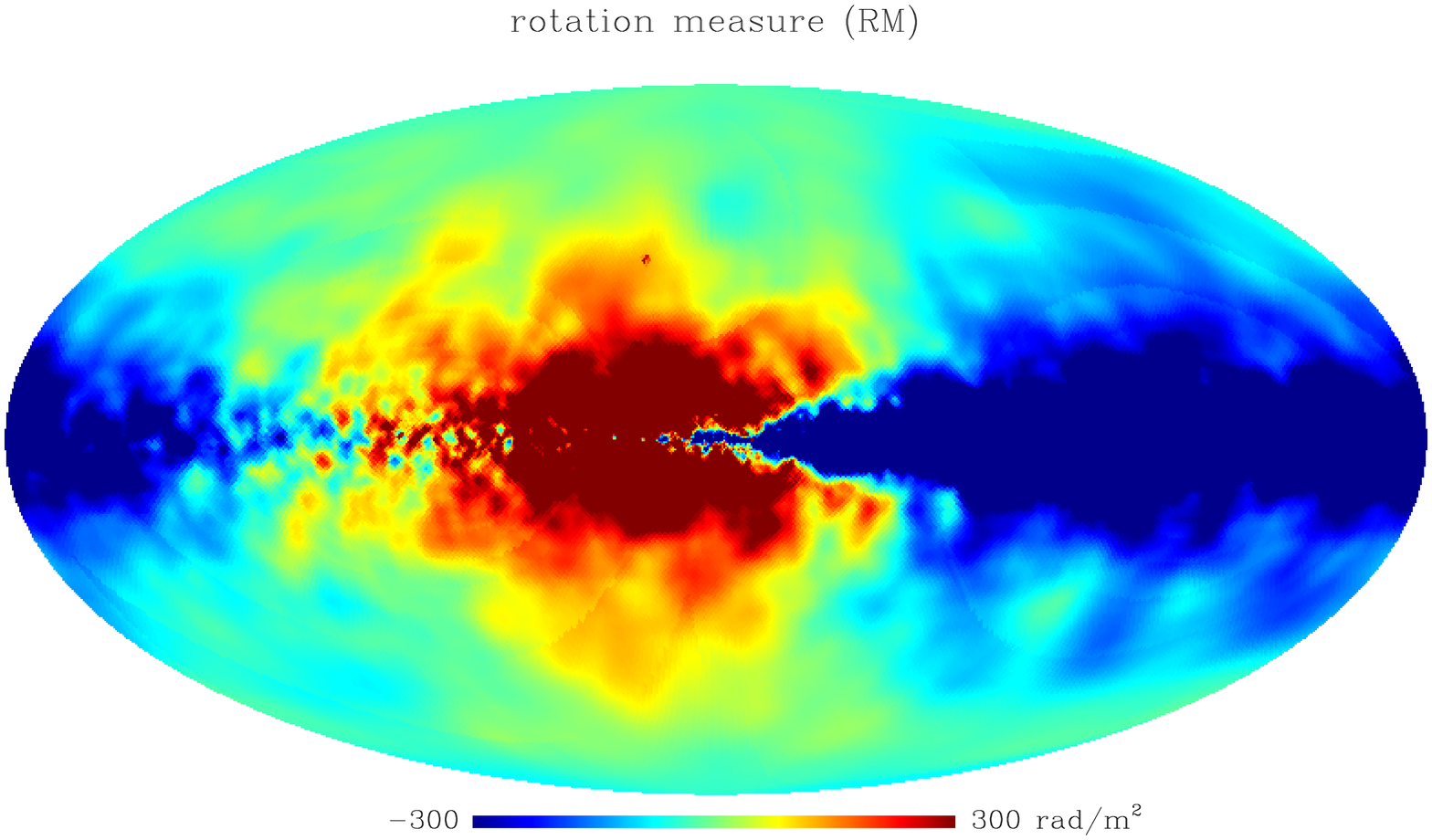}&
  \includegraphics[angle=+90, width=0.31\textwidth]{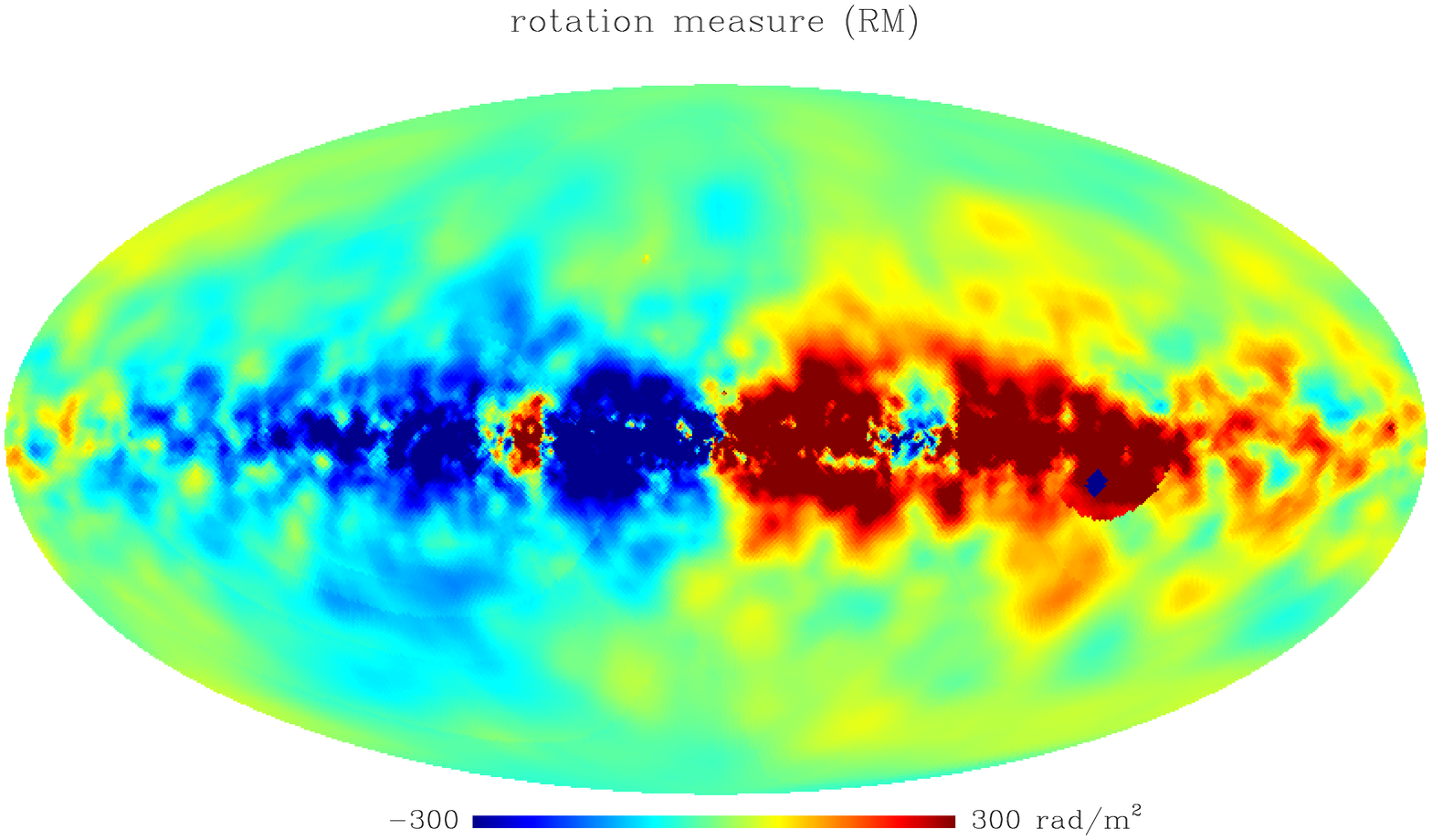}&
  \includegraphics[angle=+90, width=0.31\textwidth]{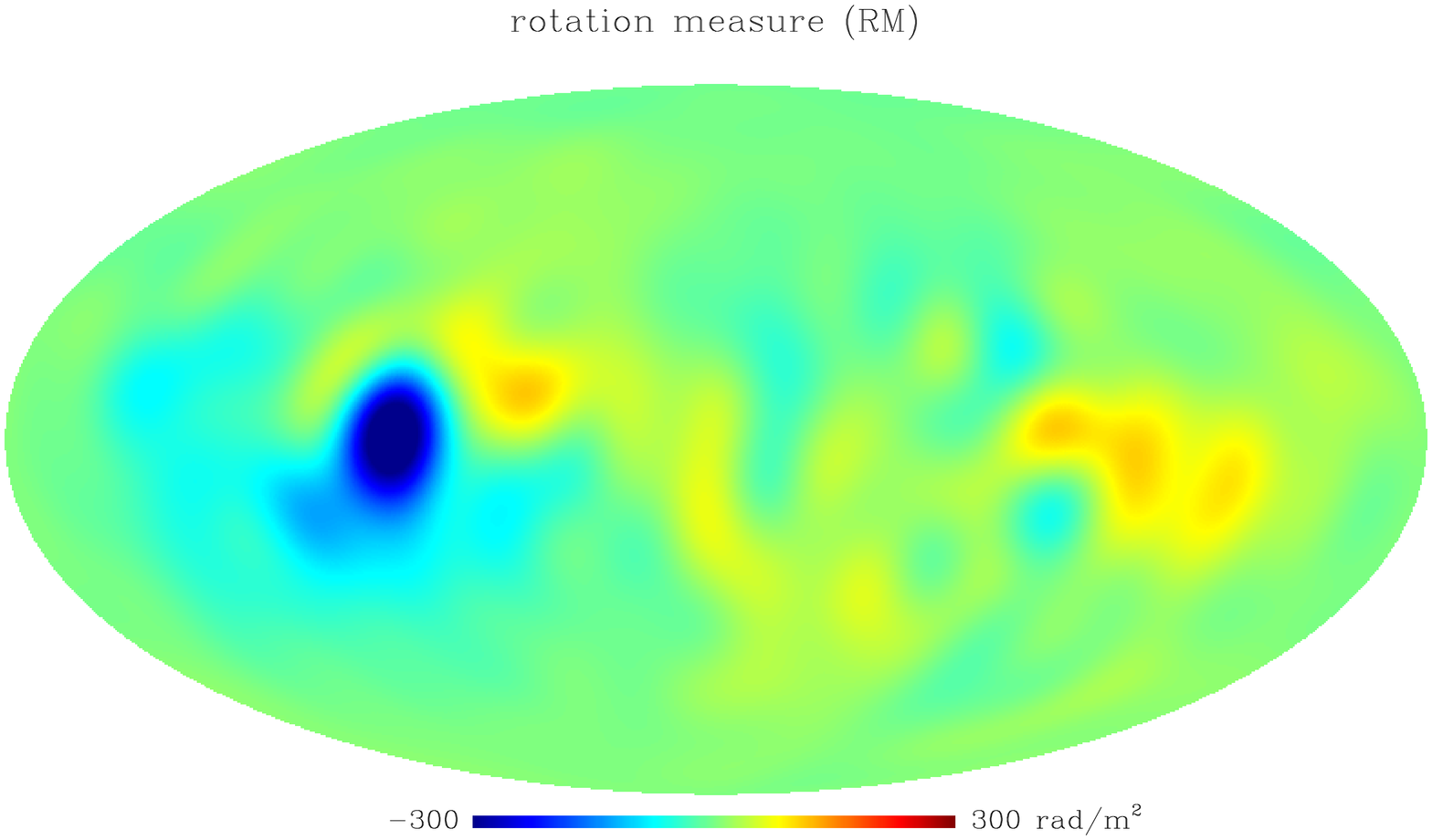}
\end{tabular}
  \caption{Comparison of mock observations based on models from
  \citet{Page2007} and \citet{Sun2008} with observed data. Each row is
  respectively total intensity I, polarized intensity P, polarization angle PA
  and rotation measure RM. The third column contains the observables. I is from
  \citet{Haslam1981, Haslam1982}, P and
  PA are WMAP observations \citep{Page2007, Hinshaw2008aph} , while the RM is
  from \citet{Dineen2005} using the \citet{Simard1981} data.} 
  \label{fig::MC}
\end{figure*}

In this section we compare our example simulation with observed data.
The mock total intensity I, polarized intensity P, polarization angle PA and RM
maps for our two different field configurations and slightly different cosmic-ray electron
distributions of \citet{Sun2008} and
\citet{Page2007} are shown in comparison to the observed synchrotron emission
at 0.408 MHz (total intensity) and 23 GHz (polarized emission), and a RM map extrapolated from the existing still sparse RM
observations (Fig. \ref{fig::MC}). All maps in this figure and throughout the
paper are at a resolution of ${\rm NSIDE}={\rm 128}$, $\Delta r = 0.21 {\rm
  kpc}$, $H_{\rm obs}=2$, $H_{\rm max}=3$ (see definitions in App. \ref{app::HPG}). A Gaussian random
field with a Kolmogorov spectrum $k^{-5/3}$ with a lower cutoff at
$k_0=1{\rm kpc^{-1}}$ \citep[see, e.g.,][]{Han2004}, and an upper cutoff
given by the Nyquist frequency of the simulation box
$k_{Nyquist}=3.2{\rm kpc^{-1}}$, was added on both large-scale GMF models. We
caution that this scale range is insufficient to describe the turbulent field
in the Galaxy, as can be seen when noticing the even smaller-scale fluctuations
present only in the observations, however it serves our
demonstration purposes. Following \citet{Sun2008}
the random field amplitude is $\langle B_{ran}^2 \rangle^{1/2} =3 {\rm \mu G}$.

Magnetic field models in the literature are mostly designed to fit a
  selected set of data. For example \citet{Page2007} fit PA observations, and, as can be seen in
Fig. \ref{fig::MC}, does this remarkably
well. However, their model evidently fails to reproduce any of the other
observations \citep[as shown by][]{Sun2008}. Furthermore \citet{Sun2008} note
that most spiral magnetic field configurations have rather similar PA,
suggesting a highly degenerate fitting problem. 

The \citet{Sun2008} models for the GMF and the cosmic-ray electron distribution
were designed to fit longitude and latitude RM profile
observations and I and P observations at 408~MHz, 1.4~GHz and 22.8~GHz, but no
optimization was performed for the PA observations. Note also that
\citet{Sun2008}, unlike \citet{Page2007}, who did a parameter scan, adjust the
model parameters based on trial and error. 
The attentive reader might notice a faint horizontal stripe like pattern seen
in the I and P simulations of this model, which is due to a combination of our limited radial step size $\Delta r = 0.21 {\rm kpc}$, and
the abrupt cut of the cosmic-ray electron density at 1 kpc away from the
Galactic plane (as discussed in Sec. \ref{sec::CREmodel}). In other words, the
code cannot resolve a sharp cut. A smaller $\Delta r$, of course, makes the
feature vanish, though it would be inconsistent with our recommendation in
App. \ref{app::HPG} to peg $\Delta r$ to the $\rm NSIDE$ parameter.
Probably the most striking difference between the two models lies in the
different RM maps, mainly due to so called field reversals in the
\citet{Sun2008} model, which are absent in the WMAP model. 

\subsection{Helicity in the ISM?}
\begin{figure}
  \centering
  \includegraphics[angle=+90, width=0.4\textwidth]{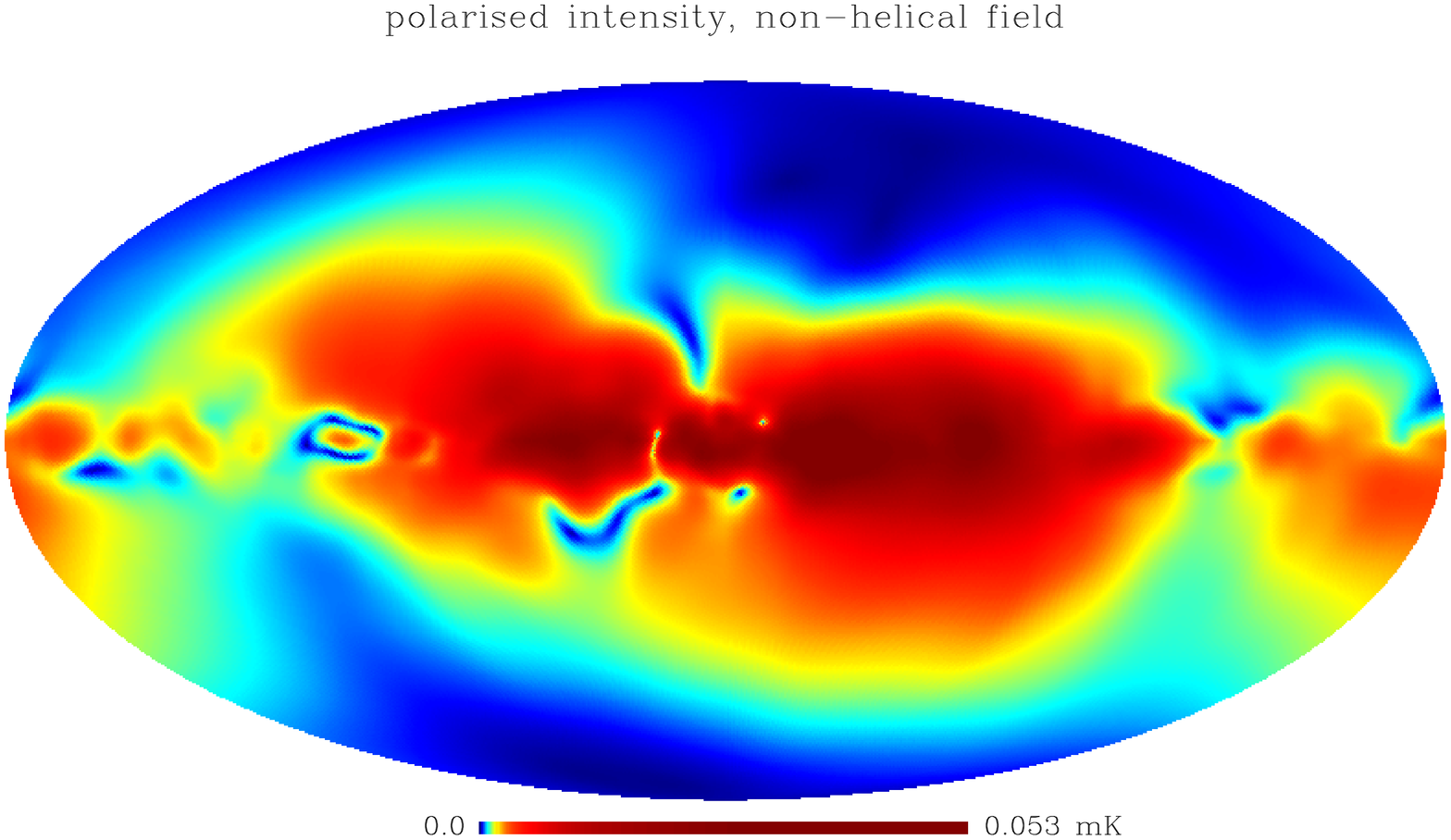}
  \includegraphics[angle=+90, width=0.4\textwidth]{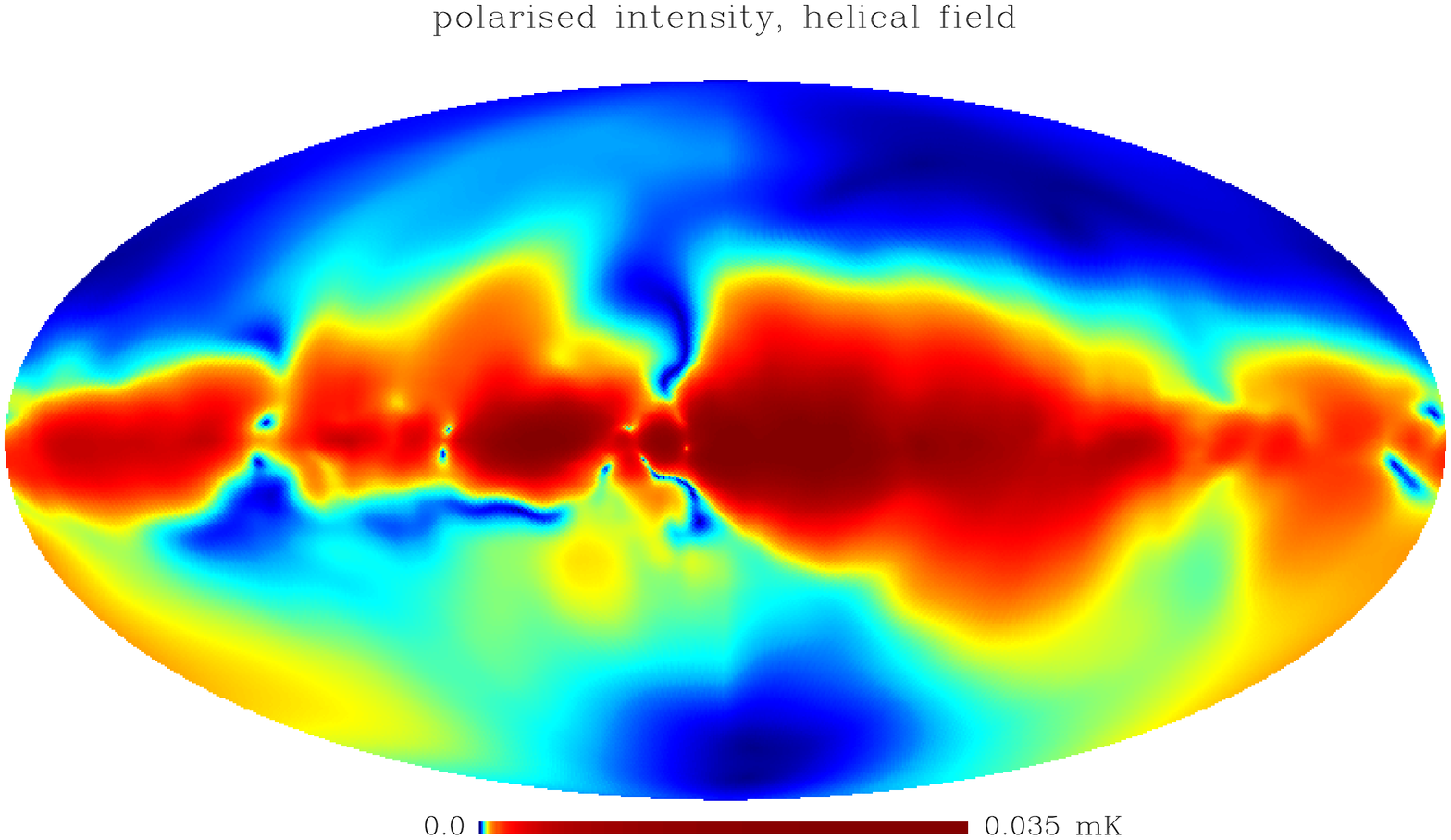}
  \caption{Polarized intensity of a field without and one with helicity. The
  colors in the plot are equally distributed over area to enhance the
  visualization of structures at the expense of proportionality information.}
  \label{fig::HF}
\end{figure}

Our code provides the possibility to study the effects of magnetic helicity on observations.
In Fig. \ref{fig::HF} a field realization with and without helicity are compared. Although
they are clearly different, there is no qualitative difference between the
single frequency maps from fields with
helicity and without. These results reinforce that measuring
helicity in the ISM is difficult and the necessary multi frequency methods to extract it
e.g. from Faraday tomography observations \citep{Brentjens2005, Schnitzeler2007} still have to be developed. The
availability of a tool for generating
mock observations might be of assistance for the development of such methods
and for feasibility studies of
observations aiming at determining the helicity in the ISM.

\subsection{UHECR deflection outputs}
\begin{figure}
  \centering
  \includegraphics[angle=+90, width=0.4\textwidth]{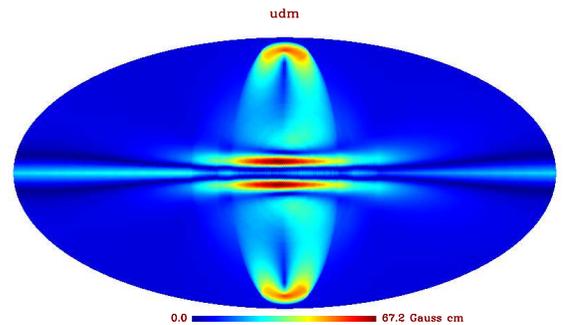}
  \caption{UHECR deflection measure map from the \citet{Sun2008} large-scale magnetic field
  model. Currently only a couple of hundred UHECR detections have been reported.}
  \label{fig::defl}
\end{figure}

\textsc{hammurabi} as a generic tool to study the GMF also helps to prepare for
UHECR-based magnetic field studies, once the sources of the UHECR particles are
identified. We included the option to compute the
deflection of ultra-relativistic charges by the GMF in the code. This is yet
another observable by which the GMF may be further constrained in the
future. Presently there are only a couple of hundred registered events and
their origin is still speculative \citep{Auger2007}.
Fig. \ref{fig::defl} represents the deflection
intensity for the large-scale \citet{Sun2008} model. Note that the strong halo field is
controversial (as discussed in Sect. \ref{sec::CREmodel}). The
deflection angle of an individual UHECR is obtained by multiplying the deflection intensity map with
$Z \, q_e/p \, c$ of this particle, as in Eq. \ref{eq::UHECRDA}.

\subsubsection{A LOFAR, SKA, WMAP, Planck tool}
The frequency range of \textsc{hammurabi} synchrotron simulations, from
$\sim 100$~ MHz up
to $\sim 100$~ GHz, covers currently running experiments like WMAP as well as the
upcoming generation of radio telescopes: Planck, LOFAR, SKA and the like.
\textsc{hammurabi} has been developed to support the scientific exploitation of
the data of these experiments and to provide synthetic observations for design studies. It is
capable of generating full- or partial-sky (not shown) maps as well as individual LOS
(see Figure \ref{fig::RMLOS}), which are useful for non diffuse measurements,
 e.g. RMs. 

\begin{figure}
  \centering
  \includegraphics[angle=+90, width=0.4\textwidth]{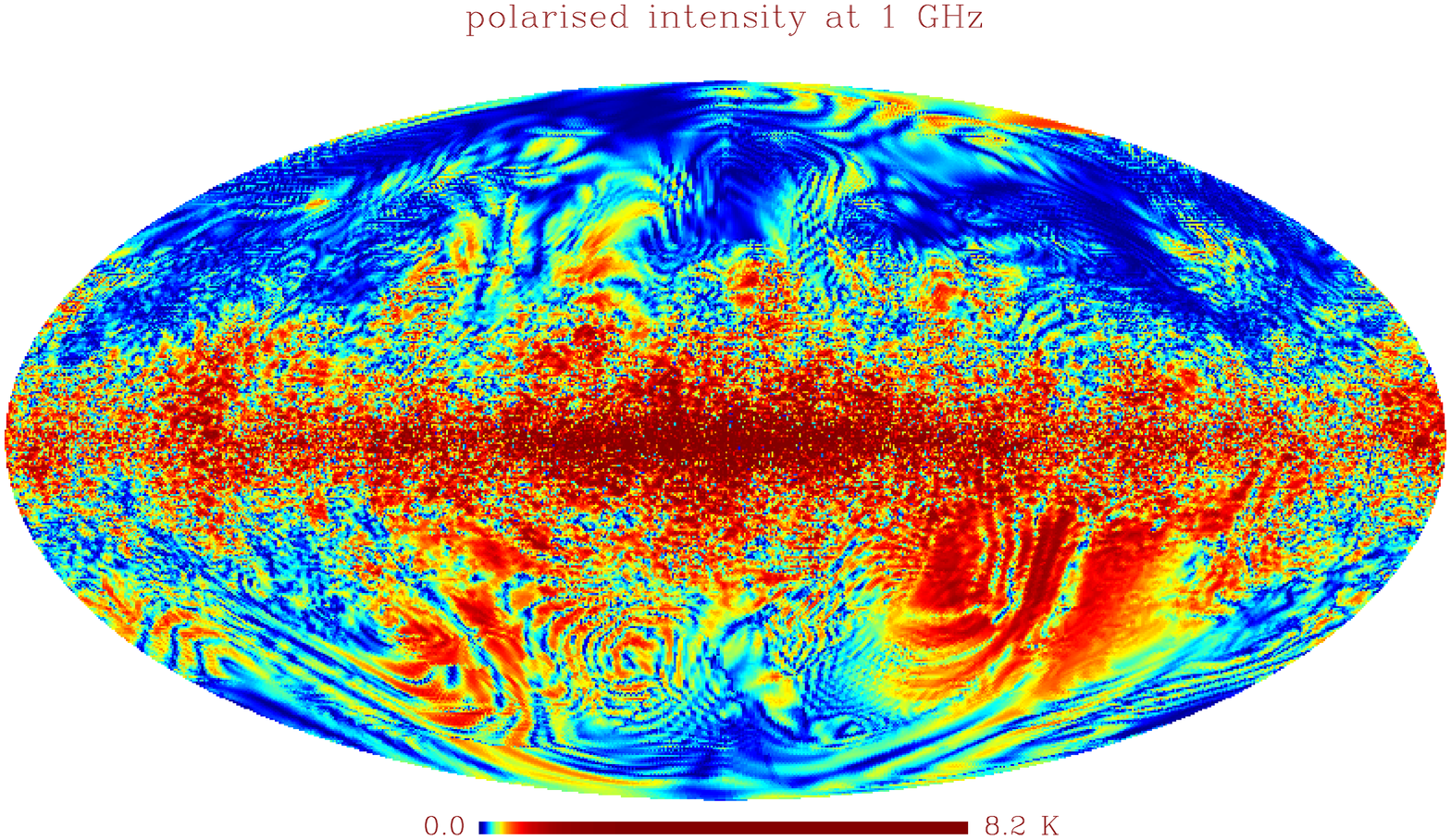}
  \includegraphics[angle=+90, width=0.4\textwidth]{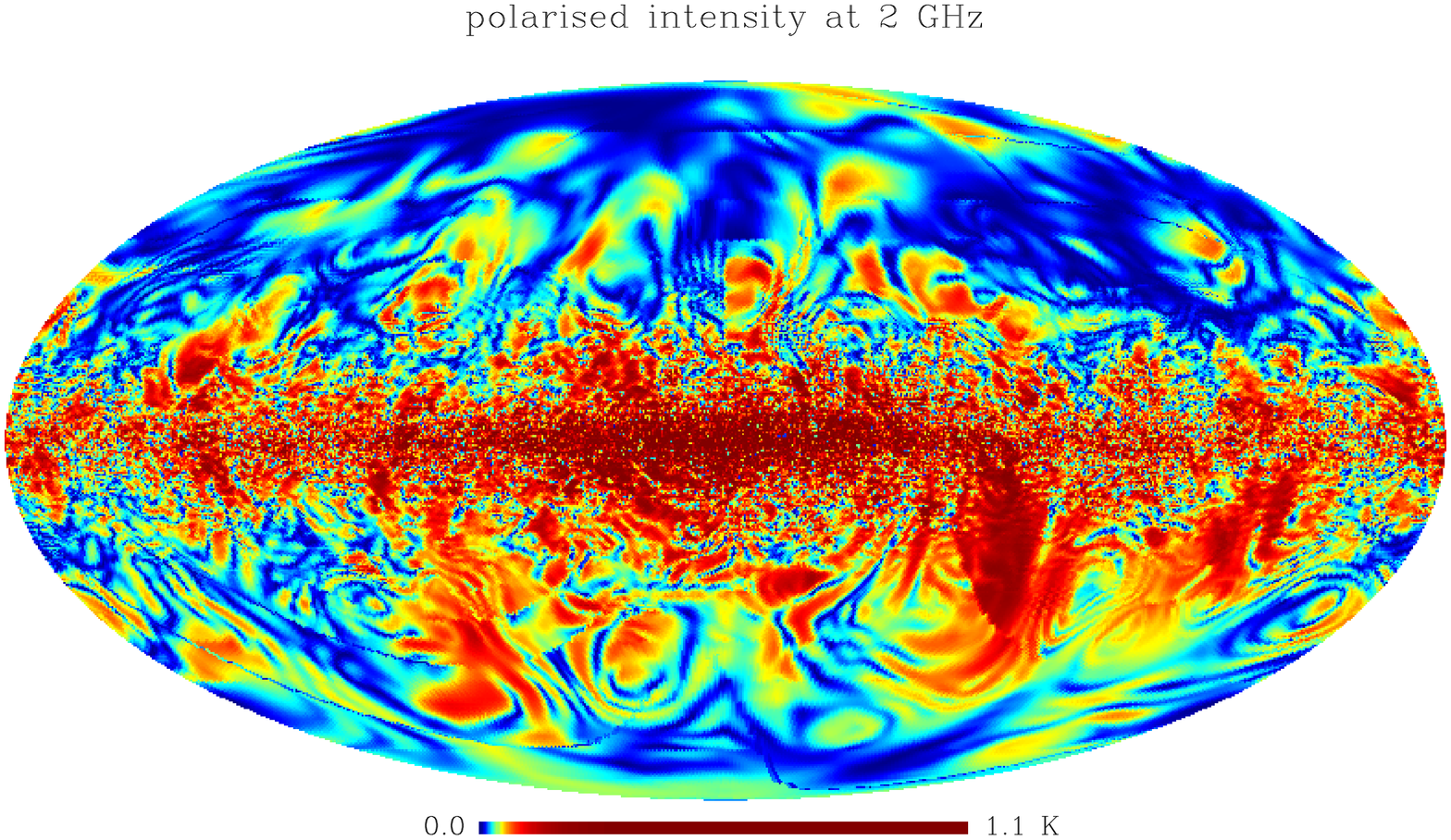}
  \caption{Simulated polarized intensity P at 1 and at 2GHz of the
  \citet{Sun2008} model. The colors in the plot are equally distributed over
  area to enhance the visualization of structures at the expense of
  proportionality information. The Faraday depolarization effect changes the
  complex patterns as a function of the wavelength. The higher thermal electron
  density on the Galactic plane keeps Faraday depolarization active long after
  it already ceased to be relevant in the Galactic halo.} 
  \label{fig::FD}
\end{figure}

\begin{figure}
  \centering
  \includegraphics[angle=+90, width=0.4\textwidth]{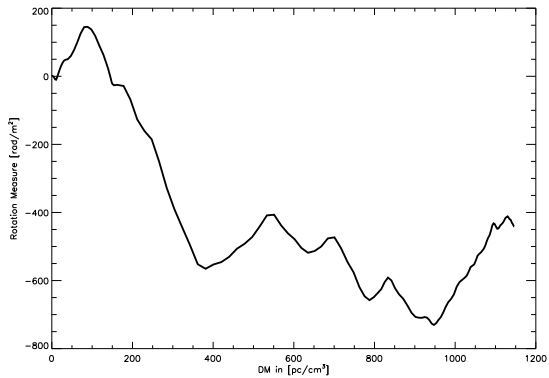}
  \caption{The simulated rotation measure as a function of dispersion
  measure. Here we used a Gaussian random field with an upper cutoff of 5 kpc and a root-mean-square of
  6 $\rm{\mu Gauss}$. Pulsar polarization measurements are, with the advent of
  SKA, expected to become very numerous, allowing us to chart the GMF in
  three-dimensions, as is shown in the present plot. } 
  \label{fig::RMLOS}

\end{figure}

Figure \ref{fig::FD} shows the full sky polarized intensity P for the
aforementioned \citet{Sun2008} model (applying their low frequency
corrections and coupling between thermal electrons and the random RM component).

In principle, multi-frequency simulations permit one to take into account
beam-width depolarization effects. Since the measurements of real telescopes
correspond to an integral over some frequency band, band-width depolarization
will also happen provided the Faraday effects are strong enough across the
band-width. Although band-width depolarization is not a standard
\textsc{hammurabi} feature, it can be emulated by coadding closely spaced
frequency maps. 

The framework of the code is built to allow additions of any other
beam-like observables such as in \citet{Waelkens2008}, \citet{Sun2008}.

\section{Conclusion}
\label{sec::Conclusion}
We have presented the \textsc{hammurabi} code, a tool for simulating rotation
measure maps, total and polarized Galactic synchrotron emission
maps taking into account Faraday rotational-depolarization effects, as well as
UHECR deflection maps.

Drawing a set of well known input models from the literature (NE2001 for the
thermal electron distribution, \citet{Sun2008} and \citet{Page2007} for the
cosmic-ray electron distribution and the Galactic magnetic field), we show
example outputs comparing them to corresponding observations to illustrate the
code's abilities as a scientific tool for charting the Galactic magnetized
plasma and the cosmic-ray electron distribution.

Full galaxy simulations are currently limited by the finite HEALPix-grid resolution, thus being unable to
probe fluctuations with characteristic lengths of less than $4 {\rm pc}$. A second caveat is that the
widely used approximation of a power-law energy distribution of the cosmic-ray
electrons has been shown to be questionable by sophisticated cosmic ray
propagation simulations \citep{Strong2007}. The degree to which this affects
the precision of our simulations has yet to be assessed.
Furthermore, our non-ray-tracing calculations of UHECR deflections might not
fulfill strict precision requirements as soon as the UHECR's trajectory through
the Galaxy is significantly bent; they serve rather as a first approximation.

It was shown that, unsurprisingly, models designed to fit only a fraction of the available
observational data on GMF might fail to reproduce the remaining observational
information not taken into account in their construction.

This is an indication that the constraints to the GMF might be highly
degenerate. Hence confronting the models with the broadest possible range of
observations on the Galactic magnetic field is paramount to achieving any
useful statement about the field.

Furthermore we have displayed \textsc{hammurabi}'s capabilities of generating
low-frequency mock observations, where Faraday effects play a significant
role. This can be used for feasibility studies and the analysis of the actual
observations of forth coming low-frequency radio telescopes like LOFAR or the
SKA (note that the latter can also observe at high-frequencies). 

\textsc{hammurabi} can be applied for constraining the Galactic magnetized plasma
and increasing our understanding on radio Galactic emission and UHECR
source locations. Already a worthy scientific goal by itself, it will also have
implications on our understanding on foreground emission subtraction for
experiments like Planck or WMAP.

\begin{acknowledgements}
AHW thanks Xiaohui Sun for remarkably good bug hunting and fruitful discussions;
Ronnie Jansson for useful comments and discussions in the development of
the code; Wolfgang Reich for fruitful discussions on radio astronomy; Andrew Strong for enlightening discussions about the
cosmic-ray electron population properties; Tony Banday for useful comments on
the manuscript; Peter Coles for providing the RM data
files; James Cordes and Joseph Lazio for kindly allowing us to make an
altered version of their NE2001 software available on the internet.  
Some of the results in this paper have been derived using the HEALPix
\citep{2005ApJ...622..759G} package. 
We acknowledge the use of the Legacy Archive for Microwave Background Data Analysis (LAMBDA). Support for LAMBDA is provided by the NASA Office of Space Science.
\end{acknowledgements}

\bibliography{hammurabi}
\bibliographystyle{aa}

\begin{appendix}
\section{The 3D HEALPix grid}
\label{app::HPG}
The \textsc{hammurabi} code generates HEALPix maps with an angular resolution
defined by the parameter ${\rm NSIDE}=2^k$, where $k \in \left [0,13
\right]$ (the maximum
$k$ is a computational limit; it could be extended only by alterations to the
original HEALPix package). 
The integration volume has a cone-like geometry. To minimize the
non-homogeneous volume sampling
induced by that, the integration volume is consecutively subdivided in so-called sub-beams. Each section of these sub-beams is contained in a shell
centered on Earth. We shall call this grid the 3D HEALPix grid. 

\begin{figure}
  \centering
  \includegraphics[bb = 660 650 1050 800,clip,  width=0.4\textwidth]{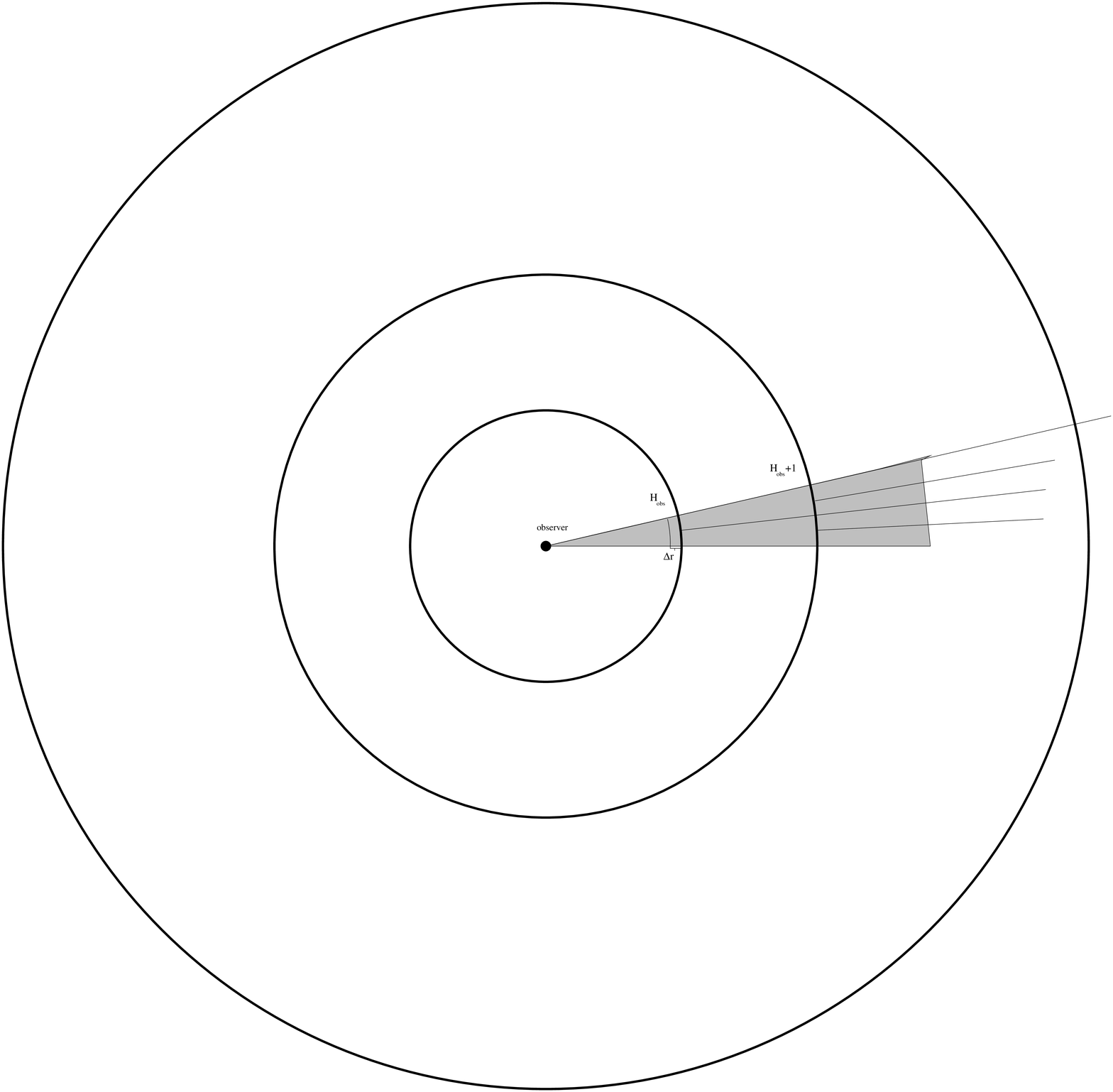}
  \caption{An example of the 3D HEALPix grid. For simplicity it is presented in
  2D. The gray shaded area corresponds to the integration volume.}
  \label{fig::shells}
\end{figure}

This subdivision will depend on 
\begin{itemize}
\item the radial length $\Delta r$ of each volume unit in the observation beam,
\item how many shells $H_{\rm max}$ in total there will be,
\item what resolution, or equivalently what ${\rm NSIDE}_{\rm obs}$ the observation shell has,
\item and finally the shell number of the observation shell $H_{\rm obs}$.
\end{itemize}
See an example in Fig. \ref{fig::shells}.

The distances at which the beam is split, or equivalently the upper boundaries
\footnote{The code will automatically round down $d_i$ to a multiple of $\Delta r$.} of the shells, are given by
\begin{eqnarray}
d_i=R_{max} 2^{i-H_{\rm max}} .
\end{eqnarray}
Where the index $i \in [1,H_{\rm max}]$.
Due to the disc-like shape of our galaxy, the effective radius out to which we
perform our computations varies and decreases sharply towards high latitudes
and towards the anti-center direction.

The maximum volume for each shell is a constant corresponding approximately to
\begin{equation}
 \label{eq::Vmax} V_{max}=4\pi d_{H_{\rm obs}}^2 \frac{\Delta r}{12 \cdot
   \mathrm {NSIDE}_{\mathrm{obs}}^2} .
\end{equation} 
We suggest \footnote{Choices of any other constant are of course possible.} a definition of $\Delta r$ such that the maximum volume unit at the largest
shell is close to a cubic form.

Note that the maximum volume unit $V_{max}$ decreases with higher observation
resolution. It defines the minimum scale which can be resolved with the code
for a certain resolution.
Fluctuations smaller than that cannot be resolved by the code. 
This is a formal upper limit, since the Galactic regions more prone to have
small scale fluctuations are likely not at the far end of the observation
volume, where $V_{max}$ is located, but instead at regions closer to the
observer, where the spatial resolution is finer.

\subsection{Line of sight integration}
\label{app::LOSI}
For the $i$-th volume element of a cone, according to the formulae in the
theoretical section \ref{sec::physics}, we compute the following, 
\begin{equation}
\label{simequation}
\left\{
\begin{array}{rcl}
I_i^{\rm syn} &=& C_I B_{i,\,\bot}^{(1-p)/2} \nu^{(1+p)/2}\Delta r\\[2mm]
P_i      &=& C_{P} B_{i,\,\bot}^{(1-p)/2} \nu^{(1+p)/2}\Delta r\\[2mm]
RM_i          &=& a_0 n_{e\,i}B_{i,\,LOS}\Delta r\\[2mm]
\chi_i        &=& \sum_{j=1}^iRM_j\lambda^2 + \chi_{i,\,0}\\[2mm] 
Q_i           &=& P_i\cos(2\chi_i)\\[2mm]
U_i           &=& P_i\sin(2\chi_i)\\[2mm]
B_{x,\,i}       &=& B_i\cos(\chi_{i,\,0}) \\[2mm]
B_{y,\,i}       &=& B_i\sin(\chi_{i,\,0}) \\[2mm]
\end{array}
\right.
\end{equation}

Note that the calculated Stokes values $U_i$ and $Q_i$ for each volume element
include the effect of foreground RM. The CR electrons are assumed to 
follow a power law with an energy spectral index $p$ (see
Sect. \ref{sec::SyncRad}). The values $C_I$ and  $C_{P}$ are dependent on the
spatialy dependent spectral index $p$ and the spatial distribution component of
the cosmic rays $C$, and can be obtained from the formulae given in section
\ref{sec::physics} and \citet{RL}. The intrinsic polarization angle
$\chi_{i,\,0}$ is locally defined as zero if the magnetic field is pointing in positive
Galactic-coordinates $l$-direction and $\pi/2$ when the magnetic field is
pointing in positive Galactic-coordinates $b$-direction.

The intensities, RMs and UDMs for a pixel are straightforward to compute. They consist
of an integral of the contributions from all the volume units as below
(Eq. \ref{eq::SC}). For the intensities, in case of a subdivision of the
observation cone into sub-cones, the computation is done
by averaging the set of sub-cones. This, however, is not done for the UDM and RM
values, since they are respectively observed as being linearly dependent on $\lambda^3$ and
$\lambda^2$, where $\lambda$ is the wavelength. The beam averaging could
destroy that linear behavior \footnote{The final RM and UDM
  value is defined as the sum of the highest possible resolution sub-beam to
  all parent beam contributions. Thus one intensity pixel might have several
  corresponding RM and UDM pixels, since those are computed for higher resolutions,
  while the former averages over the same higher resolution computations.}. 
The UHECR deflection measure is given by $UDM = \sqrt
{UDM_x^2+UDM_y^2}$, while the deflection orientation is given by $\Theta_{\rm
  defl} = \arctan \frac{UDM_x}{UDM_y}$.
\begin{equation}
\label{eq::SC}
\left\{
\begin{array}{ccc}
I           & = & \sum_iI_i\\[2mm]
Q           & = & \sum_iQ_i\\[2mm]
U           & = & \sum_iU_i\\[2mm]
RM          & = & \sum_iRM_i\\[2mm]
UDM_{x} & = & \sum_iB_{x,\,i} \Delta r \\[2mm]
UDM_{y} & = & \sum_iB_{x,\,i} \Delta r \\[2mm]
\end{array}
\right.
\end{equation} 

\section{Magnetic field models}
\label{app::MFM}
For the convenience of the reader we present the parameterizations of the
galactic magnetic field folowing \citet{Page2007} and
\citet{Sun2008}. Everything is in the usual cylindrical
coordinates, whose origin lies at the Galactic center. The x-axis of the
coordinate system points in the opposite direction of the Sun, while the z-axis
points towards the Galactic north.
\begin{itemize}
\item \citet{Page2007} writes
\begin{eqnarray}
\vec{B}(r,\phi,z) & = & B_0 \left [ \cos \psi(r) \, \cos \chi(z)\hat{r} +
   \right . \\ 
&& \left . \sin \psi(r) \,  \cos \chi(z)\hat{\phi} + \sin \chi(z)\hat{z}
   \right ]
\end{eqnarray}

Here $\psi(r) = \psi_0+\psi_1 ln(r/8 {\rm kpc})$, $\chi(z) = \chi_0 \,
\tanh(z/1 {\rm kpc})$. The radial variable $r \in [{\rm 3 kpc}, {\rm 8 kpc}]$,
$\chi_0={\rm 25^o}$, $\psi_1={\rm 0.9^o}$ and $\psi_0 = {\rm 27^o}$. $B_0$ is
not specified in \citet{Page2007} and we put $B_0=4 \, \mu G$.

\item While \citet{Sun2008} presents some suggestions for fields,
  and we pick their {\it ASS+RING} parameterization:
\begin{equation}
\left\{
\begin{array}{rcl}
B^D_{\hat r}    & = & D_1(R,\phi,z)D_2(R,\phi,z)\sin p\\[2mm]
B^D_{\hat \phi} & = &-D_1(R,\phi,z)D_2(R,\phi,z)\cos p\\[2mm]
B^D_{\hat z}    & = & 0 
\end{array}
\right.
\end{equation}
Where
\begin{equation}\label{ring}
D_1(r,z)=\left\{
\begin{array}{cl}
\displaystyle{B_0\exp\left(-\frac{r-R_\odot}{R_0}-\frac{|z|}{z_0}\right)} & 
r>R_c\\[4mm]
B_c & r \leq R_c
\end{array}
\right.
\end{equation}
Here $R_0=10$~kpc, $z_0=1$~kpc, $R_c=5$~kpc, $B_0=2$~$\mu$G and
$B_c=2$~$\mu$G.
and 
\begin{equation}
D_2(r)=\left\{
\begin{array}{ll}
+1 & r>7.5\,\,{\rm kpc}\\
-1 & 6\,\,{\rm kpc}<r\leq 7.5\,\,{\rm kpc}\\
+1 & 5\,\,{\rm kpc}<r\leq 6\,\,{\rm kpc}\\
-1 & r\leq 5\,\,{\rm kpc} 
\end{array}
\right.
\end{equation}
While the halo field is given by

\begin{equation}
B^H_\phi(r,z)=B^H_0\frac{1}{1+\left(\displaystyle{\frac{|z|-z^H_0}{z^H_1}}\right)^2}
\frac{r}{R^H_0}\exp\left(-\frac{r-R^H_0}{R^H_0}\right)
\end{equation}

and the parameters are $z^H_0=1.5$~kpc, $B^H_0=10$~$\mu$G, $R^H_0=4$~kpc,
$z^H_1=0.2$~kpc (for $|z|<z^H_0$), and $z^H_1=0.4$~kpc (otherwise).
\end{itemize}

\end{appendix}

\end{document}